\documentclass[journal,draftcls,onecolumn,12pt,twoside]{IEEEtranTCOM}
\normalsize

\usepackage{cite}
\usepackage{amsmath,amssymb,amsfonts}
\usepackage{bm,tikz,pgfplots,stmaryrd,wrapfig,siunitx}
\usepackage{algorithmic}
\usepackage{graphicx}
\usepackage{textcomp}

\usepackage{xcolor}
\usepackage{lipsum}  
\usetikzlibrary{positioning,arrows,shapes.geometric,calc}
\usetikzlibrary{decorations.markings}
\tikzstyle{block} = [draw, rectangle, minimum height=1em, minimum width=2em,text centered]
\tikzstyle{lblock} = [draw, rectangle, minimum height=2em, minimum width=4em,text centered, text width=5em]
\tikzstyle{l2block} = [draw, rectangle, minimum height=1em, minimum width=4em,text centered, text width=9em]
\tikzstyle{l3block} = [draw, rectangle, minimum height=1em, minimum width=4em,text centered, text width=11em]
\tikzstyle{ccmblock} = [draw, rectangle, minimum height=2em, minimum width=2em,text centered, text width=4em]
\tikzstyle{llblock} = [draw, rectangle, minimum height=30em, minimum width=2em,text centered]
\tikzstyle{summer} = [draw, circle, text centered]
\tikzstyle{attenuatorR} = [draw, regular polygon, regular polygon sides=3,shape border rotate=30, text centered]
\tikzstyle{antenna} = [draw, regular polygon, regular polygon sides=3,shape border rotate=180, text centered]
\def\BibTeX{{\rm B\kern-.05em{\sc i\kern-.025em b}\kern-.08em
    T\kern-.1667em\lower.7ex\hbox{E}\kern-.125emX}}

\tikzstyle{vecArrow} = [thick, decoration={markings,mark=at position
   1 with {\arrow[semithick]{open triangle 60}}},
   double distance=1.4pt, shorten >= 5.5pt,
   preaction = {decorate},
   postaction = {draw,line width=1.4pt, white,shorten >= 4.5pt}]
\ifCLASSINFOpdf
\else
\fi

\begin{document}
\title{An Efficient Slow-Time Adaptation for Massive MIMO Hybrid Beamforming in mm-Wave Time-Varying Channels}

\author{Anil~Kurt,~\IEEEmembership{Student Member,~IEEE,}
        Gokhan~Muzaffer~Guvensen,~\IEEEmembership{Member,~IEEE}% <-this % stops a space
\thanks{The authors are with the Department
of Electrical and Electronics Engineering, Middle East Technical University (METU), Ankara, TURKEY (e-mail: anilkurt@metu.edu.tr; guvensen@metu.edu.tr)}}

\maketitle

\begin{abstract}
%\boldmath
%In this paper, adaptive hybrid beamforming methods are proposed for millimeter-wave range massive MIMO systems considering single carrier wideband transmission in uplink data mode. A statistical analog beamformer is adaptively constructed in slow-time, while the channel is time-varying and erroneously estimated. Proposed recursive filtering approach is shown to bring a remarkable robustness against estimation errors when instantaneous effective channel is assumed to be known. Approximated expressions are obtained for channel covariance matrices that decouple angular spread and center angle of multipath components. With these expressions, modified adaptive construction methods  use only the estimated power levels on angular patches and they are shown to be very efficient such that they reduce computational complexity significantly while the performance remains almost the same. Finally, it is shown that proposed methods keep their advantages when estimated instantaneous effective channel information is utilized.

In this paper, adaptive hybrid beamforming methods are proposed for millimeter-wave range massive multiple-input-multiple-output (MIMO) systems considering single carrier wideband transmission in uplink data mode. A statistical analog beamformer is adaptively constructed in slow-time, while the channel is time-varying and erroneously estimated. A recursive filtering approach is proposed, which aims robustness against estimation errors for generalized eigen-beamformer (GEB). Approximated expressions are obtained for channel covariance matrices that decouple angular spread and center angle of multipath components. With these expressions, modified adaptive construction methods for GEB are proposed, which use only the quantized estimated power levels on angular patches. The performances of the proposed slow-time adaptation techniques for statistical Massive MIMO beamforming are evaluated in terms of the output signal-to-interference-and-noise-ratio (SINR), instantaneous channel estimation and beam accuracy. They are shown to be very efficient
such that the computational complexity is significantly reduced while the performance remains almost the same as that of the ideal GEB even in large angular estimation errors.
\end{abstract}


% \fi
\IEEEpeerreviewmaketitle

\section{Introduction}

\IEEEPARstart{M}{illimeter-wave} (mm-wave) range massive multiple-input multiple-output (MIMO) systems are expected to take part in 5G beyond systems, to enable much larger data rates \cite{Andrews14}. Thanks to mm-wave conditions, the limiting factor on the number of antennas is not size, but the cost and energy of employing radio frequency (RF) chains for each antenna element. A solution is to use two-stage-processing with hybrid beamformer structure, which reduces the number of RF chains together with the number of coefficients to be estimated for channel state information (CSI) \cite{Molisch17}. Joint Spatial Division and Multiplexing (JSDM) is proposed in literature to accomplish two-stage-processing \cite{Adhikary14}. According to this idea, the first processing stage, analog beamformer, is constructed using long-term statistics and reduces the signal dimension with the help of an efficient grouping of users \cite{Chen16}. The second processing stage, digital beamformer, operates in fast-time with reduced number of channel coefficients. 

\subsection{Related Literature}
Mobility and tracking concepts in the context of mm-wave massive MIMO draw interest in literature recently. System design and signalling scheme play a significant role in the perception of these terms, and determine the importance devoted. These determining factors are beamformer structure, the philosophy behind the use of hybrid beamformer if used, multiplexing preference in multiuser cases, channel estimation scheme, and frequency-selectiveness of the channel (hence the preference of the signalling, such as single-carrier (SC) or orthogonal frequency division multiplexing (OFDM)). 

User mobility in fully analog beamforming is mainly about tracking angle of arrival (AoA). Movement is modeled and generally Kalman filter (and its variants) is implemented to track AoA \cite{Zhang16,Va16,Talvitie18,Liu19,Shaham19}. Then beamformer is constructed as the related (or nearby if constrained) steering vector, due to the phase-only beamforming assumption. For fully digital beamforming, \cite{Jayaprakasam17} jointly uses erroneous AoA estimate from Kalman filter and training sequence to optimize beamforming weights. \cite{Larew19} designs low overhead sounding beams via unscented Kalman filtering. 

Usually, hybrid beamforming is used merely as a solution for physical limitations on the number of RF chains, in the pursuit of making use of large antenna arrays to cope with the expectations from 5G and beyond systems. Consequently, studies focus on the hybrid-beamforming adaptation of widely accepted fully-digital beamforming assumptions such as OFDM usage, frequency-flat massive MIMO channel, full dimensional instantaneous channel estimation and singular value decomposition (SVD) based beamforming using most dominant singular vectors. This adaptation includes decomposition of an optimal beamformer into a phase and magnitude constrained analog beamformer and an unconstrained digital beamformer \cite{Alkhateeb14,Ghauch16,Noh17,DeDonno17,Zhu19}. The calculation of optimal beamformer generally needs the knowledge of full-dimensional channel matrix, whose estimation is problematic due to the fact that there is no access to signals at antennas in hybrid beamforming \cite{Ghauch16}. In this frame, solution to mobility of users boils down to tracking AoA of users. \cite{Zhao17} tracks AoA of users via unscented Kalman filter (UKF) and constructs the analog beamformer as steering vectors to the related angles. \cite{Gao17} studies on a fixed discrete lens array (DLA) as analog beamformer, and decreases the channel estimation time by limiting the angular sectors to be used according to the AoA estimate and movement model. \cite{Long20} implements broad learning to construct a codebook for constrained analog beamforming, adopting to the user position, to be used in approximating the left singular vectors of the channel matrix as hybrid beamformer. 

On the other hand, \cite{Haghighatshoar18} proposes two-stage second-order-statistics-based processing to exploit the advantages of JSDM framework. Its beamformer construction does not include any constraints on weight modulus, which can be perceived as hybrid beamforming with unconstrained-modulus analog beamformer, with controllable amplifiers. A single mobile user communicates with BS through a frequency flat channel, and the first processing stage uses eigenmodes of channel covariance matrices (CCMs). To cope with the mobility, CCMs are filtered in time.

Most of the existing research for massive MIMO systems adopt flat-fading channel model by considering the use of OFDM \cite{Lu14}. However, due to the drawbacks of OFDM transmission (e.g., high peak-to-average-power ratio (PAPR)), the use of SC in massive MIMO systems employing mm-wave bands, exhibiting sparsity both in angle and delay plane, was considered in \cite{Buzzi18,Pitarokoilis12,Song19,Guvensen16,kurt19}. In these studies, the mitigation of inter-symbol interference (ISI) via reduced complexity beamspace processing (rather than temporal processing) motivates the use of SC in spatially correlated wideband massive MIMO systems \cite{Song19,Guvensen16,kurt19}.

Angular sparsity in mm-wave massive MIMO channels brings the spatial domain to the solution of  multiple access problem, as a third alternative besides the time and frequency domains. It is inefficient not to utilize spatial division in angular sparsity. \cite{Qin18, Gao17, Zhao17, Buzzi19} study multiuser cases in time-varying environments using analog beamformers with constant-modulus weights (generally in the form of steering vectors or columns of DFT matrix). However, since analog beamformer is constrained, interference rejection ability is limited. Further, it is seen that locations or channel parameters of interferers do not play a significant role in the calculation of analog beamformer weights. Therefore, mobility of interfering users does not constitute a special case in this context. However, it is shown in \cite{kurt19} that constrained analog beamformer (DFT beamformer) is vulnerable against near-far effect. 

To sum up, weights of analog and digital stages of hybrid beamformer are generally calculated via SVD of instantaneous channel matrix in fast-time. Under constraints, analog beamformer is not efficiently matched to the slowly-varying parameters of the channel (angular properties of intended and interfering users, and multipath components in frequency-selective channels). It brings the need of fast-time update of analog beamformer, full-dimensional estimate for instantaneous channel, and inefficient use of spatial sparsity for multiple access.

\subsection{Contributions}

This work considers a mm-wave massive MIMO system with hybrid analog/digital beamforming, serving multiple mobile users and utilizing wideband SC communication which experiences a frequency selective channel. The main motivation in this paper is to propose adaptive beamforming solutions in a time-varying channel where multiple users are moving. These solutions are built upon the system detailed in our previous work \cite{kurt19} where user positions are stationary. In \cite{kurt19}, a second-order-statistics-based analog beamformer named generalized eigen-beamformer (GEB) is proposed. Its weights are assumed to be unconstrained, but it is employed only in base station (BS) and needs slow-time updates. In return, it is shown to be very beneficial in terms of spectral efficiency due to its powerful interference cancelling ability with broad nulls. This ability is exploited in spatial multiplexing between users, and ISI-mitigation via spatial equalization. In addition, since it governs slowly changing parameters of the channel such as AoA, instantaneous channel estimation can be done in reduced-dimensional well-constructed eigenspaces, resulting in a remarkable ease in estimation, exhibiting success even with 2-3 training symbols. On the other hand, it is shown that it is vulnerable to AoA estimation errors but tolerant to angular spread (AS) errors.

This paper focuses on slow-time adaptation of GEB to the mobility of users, which is computed using generalized eigendecomposition of signal and interference CCMs. That is, it is not a basic function of AoA of intended user. To exploit the correlation of AoA in time, recursive filtering of CCMs are offered, which exhibits a remarkable robustness against estimation errors. It means that aforementioned intolerance of GEB to AoA errors is overcome, further relaxing requirements from an AoA and AS estimator. Also, in order to bring ease to the analytical construction of second order statistics, namely CCMs, for simpler interaction between outputs of an angular estimator and beamformer adaptation, alternative ways of CCM construction is proposed and expressed as analytical structures. Although they are applied on GEB in this paper, these analytical structures for CCMs can also be used for any statistical beamformer design method. Together with a modification on GEB, new adaptive beamformer construction methods are proposed, namely Wiener filter type and whitening filter type adaptive analog beamformers. They show remarkable decrease in the computational complexity without losing from the performance. Moreover, proposed methods are shown to maintain aforesaid advantages against mobility when perfect CSI assumption is removed and an instantaneous channel estimator is cooperating. 

To sum up, main contributions of this paper are 
\begin{itemize}
\item A filtering scheme of CCMs to exploit the correlation of AoA in time against user mobility and AoA estimation errors,
\item A beamformer adaptation scheme where main-lobe and multiple broad nulls are steered simultaneously in a nearly-optimal manner,
\item Approximate analytical expression for CCMs, which decomposes AoA and AS informations, together with complexity-reducing derivations such as angular patching and power level quantization, which can be used in different statistical beamforming applications,
\item Low-complexity adaptive beamforming construction methods which are invented by utilizing aforementioned analytical derivations on CCMs,
\item Assessment of cooperation of adaptive statistical beamforming and instantaneous channel estimation in terms of output signal-to-interference-and-noise-ratio (SINR) and normalized mean squared error (MSE), which exhibits nearly-optimal performance even in large angular pointing errors.
\end{itemize}

\section{System Model}
We consider a massive MIMO system operating in mm-wave bands and employing SC where a BS with $N$ antennas serves single-antenna user terminals (UTs) in time-domain duplex (TDD) mode. UTs are partitioned into $G$ groups such that angularly adjacent UTs are in the same group. The $g$\textsuperscript{th} group consists of $K_g$ users. Each group has $M^{(g)}$ resolvable multipath components (MPCs) arriving at BS with different AoA and delays. UTs in the same group share properties of these MPCs. User grouping idea in JSDM framework is utilized.

%\begin{figure}[htbp]
%\centerline{\includegraphics[width=0.3\textwidth]{Scenario.png}}
%\caption{User grouping and MPCs ($G=2$, $K_1=3$, $K_2=2$ and $L=2$)}
%\label{scenario}
%\end{figure}
\subsection{Wideband SC Transmission}\label{wsctr}
 We consider the BS has hybrid beamformer configuration. Although signals are analog until the digital beamformer stage in such a system, we express their equivalent digital forms according to Nyquist theorem  assuming sampling rate after pulse matched filtering is not less than bandwidth. The observation vector $\mathbf{y}_n$ of size $N$ consisting of samples from signals received by antenna elements in uplink data mode is 
 %\begin{equation} \label{gen_full1} \mathbf{y}_n= \sum_{g=1}^{G} \sum_{k=1}^{K_g} \sum_{l=0}^{L-1} \mathbf{h}_l^{(g_k)} x_{n-l}^{(g_k)} + \mathbf{n}_n \end{equation} or 
 \begin{equation} \label{gen_full2}
\mathbf{y}_n= \sum_{g=1}^{G} \sum_{l=0}^{L-1} \mathbf{H}_l^{(g)}\mathbf{x}_{n-l}^{(g)} + \mathbf{n}_n ,
\end{equation} where $\mathbf{H}_l^{(g)}\triangleq\left[\mathbf{h}_l^{(g_1)} , \mathbf{h}_l^{(g_2)} , ... , \mathbf{h}_l^{(g_{K_g})} \right]$ is channel matrix of size $N \times K_g$ and $\mathbf{h}_l^{(g_k)}$ are channel vectors. $\mathbf{x}_n^{(g)}\triangleq\left[x_n^{(g_1)} , x_n^{(g_2)} , ... , x_n^{(g_{K_g})} \right]^{T}$ is symbol vector of size $K_g$ and $x_n^{(g_k)}$ are transmitted symbols. $\mathbf{n}_n$ is the additive white Gaussian noise (AWGN) vector. Subscripts and superscripts $(\cdot)^{(g)}$, $(\cdot)^{(g_k)}$, $(\cdot)_l$ and $(\cdot)_n$ indicate affiliations to the group $g$, $k$\textsuperscript{th} user of the group $g$, $l$\textsuperscript{th} delay and $n$\textsuperscript{th} time instance, respectively. The $m$\textsuperscript{th} MPC of $g$\textsuperscript{th} group has a delay of $l=\mathcal{L}^{(g)}(m)$. Due to wideband transmission and temporal sparsity of mm-wave signals, $\mathbf{H}_l^{(g)}$ is nonzero for very few values of delay $l$.

Each user has statistically independent channels with independent MPCs. However, users in the same group have identically distributed channels
%\begin{equation}
%E\left\lbrace \mathbf{h}_l^{(g_k)} \left[\mathbf{h}_{l'}^{(g_{k'}')}\right]^{H} \right\rbrace  = \mathbf{R}_l^{(g)}\delta_{gg'}\delta_{kk'}\delta_{ll'}
%\end{equation} 
where $ \mathbf{h}_l^{(g_k)} \sim \mathcal{CN} \left( \mathbf{0} , \mathbf{R}_l^{(g)} \right)$ \cite{Le20}. CCM $\mathbf{R}_l^{(g)}$ depends on angular power profile of related group of users and can be obtained as \begin{equation} \label{R_gen_int}
\mathbf{R}_l^{(g)} = \int_{-\pi}^{\pi} {\rho}_{\phi,l}^{(g)}(\phi) \mathbf{u}(\phi) {\mathbf{u}(\phi)}^{H} d\phi ,
\end{equation} where $\mathbf{u}(\phi) \triangleq \frac{1}{\sqrt{N}}\left[1 \, e^{j \pi \sin(\phi)} \dots e^{j(N-1) \pi \sin(\phi)} \right]^{T}$ is the array response vector to an incidence from angle $\phi$ with respect to the perpendicular to the array, for a uniform linear array with half wavelength distance between antenna elements. ${\rho}_{\phi,l}^{(g)}(\phi)$ is the angular power profile of $l$\textsuperscript{th}-delay MPC of group $g$ and assumed to be nonzero in a very narrow angular interval considering mm-wave conditions. Also, $\rho_{\phi,l}^{(g)}(\phi)$ adjusts the power distribution among MPCs of a group in such a way that $\sum_{l=0}^{L-1} \text{tr} \left( \mathbf{R}_l^{(g)} \right) = 1$ for $g=1 , \dots , G$.

The AWGN vector $\mathbf{n}_n$ is i.i.d. for different time instances and $ \mathbf{n}_n \sim \mathcal{CN} \left( \mathbf{0} , N_0 \mathbf{I}_N \right)$ where $N_0$ is the noise power and $\mathbf{I}_N$ is the identity matrix of size $N \times N$. The transmitted symbols are assumed to be complex, zero mean, and uncorrelated: $E\left\lbrace x_n^{(g_k)} \left[x_{n'}^{(g_{k'}')}\right]^* \right\rbrace = E_s^{(g)}\delta_{gg'}\delta_{kk'}\delta_{nn'}$.
%\begin{equation} \label{symbol_cov}
% E\left\lbrace x_n^{(g_k)} \left[x_{n'}^{(g_{k'}')}\right]^* \right\rbrace = E_s^{(g)}\delta_{gg'}\delta_{kk'}\delta_{nn'}
%\end{equation}
%$E_s^{(g)}$ in (\ref{symbol_cov}) 
$ E_s^{(g)}$ is the symbol energy of $g$\textsuperscript{th} group and it is used to control the relative power distances among groups and introduce the near-far effect. $E_s^{(g)} / N_0$ gives signal-to-noise ratio (SNR) for the group $g$. Finally, observation correlation matrix $\mathbf{R}_y\triangleq E[\mathbf{y}_n\mathbf{y}_n^{H}]$ and interference-and-noise correlation matrix $\mathbf{R}_{\eta}^{(g)}$ are
\begin{align}\label{Ry}
\mathbf{R}_y & = \sum_{g=1}^G K_{g}E_s^{(g)} \sum_{l=0}^{L-1}\mathbf{R}_{l}^{(g)} + N_0 \mathbf{I}_N
\\
\mathbf{R}_{\eta}^{(g)} & = \mathbf{R}_y - K_{g}E_s^{(g)} \sum_{l=0}^{L-1}\mathbf{R}_{l}^{(g)} \label{R_eta}
\end{align}

\subsection{Mobile Channel Model}
Channel mobility is modeled as a change in center angle of each MPC $\mu_{\phi,l}^{(g)}$ with time, while angular spread $\sigma_{\phi,l}^{(g)}$ is kept constant. The center angle in the $n$\textsuperscript{th} slow-time instance (comparable with the packet duration and not to be confused with the time index in (\ref{gen_full2})) is
\begin{equation}
\mu_{\phi,l}^{(g)}[n]= \mu_{\phi,l}^{(g)}[0]+ \Delta\mu_{\phi,l}^{(g)}[n],
\end{equation}
where the mobility addend $\Delta\mu_{\phi,l}^{(g)}[n]$ is a stochastic process obtained by autoregressive first order (AR(1)) Markov process, shown as
\begin{equation}
\Delta\mu_{\phi,l}^{(g)}[n]=\alpha \Delta\mu_{\phi,l}^{(g)}[n-1] + \sqrt{1-\alpha^2} \, v[n] \text{ for } n=1,2,\dots
\end{equation} where $\Delta\mu_{\phi,l}^{(g)}[0]=0$ and $0<\alpha<1$. In addition, $v[n] \sim \mathcal{N} \left( 0 , \sigma_v^2 \right)$ and i.i.d. for different $n$. The process $\mu_{\phi,l}^{(g)}[n]$ has the mean  $\mu_{\phi,l}^{(g)}[0]$ for $n \ge 0$, and a variance converging to $\sigma_v^2$ as $n$ goes to infinity.

The AoA estimation $\hat{\mu}_{\phi,l}^{(g)}[n]$ is modeled as 
\begin{equation}
\hat{\mu}_{\phi,l}^{(g)}[n]=\mu_{\phi,l}^{(g)}[n] + e[n]
\end{equation}
where $e[n]$ is normally distributed, zero-mean and i.i.d. error term whose variance is $\sigma_{est}^2$.

\subsection{Hybrid Beamformer Structure for Uplink}
\subsubsection{Receiver Processing}\label{recproc}

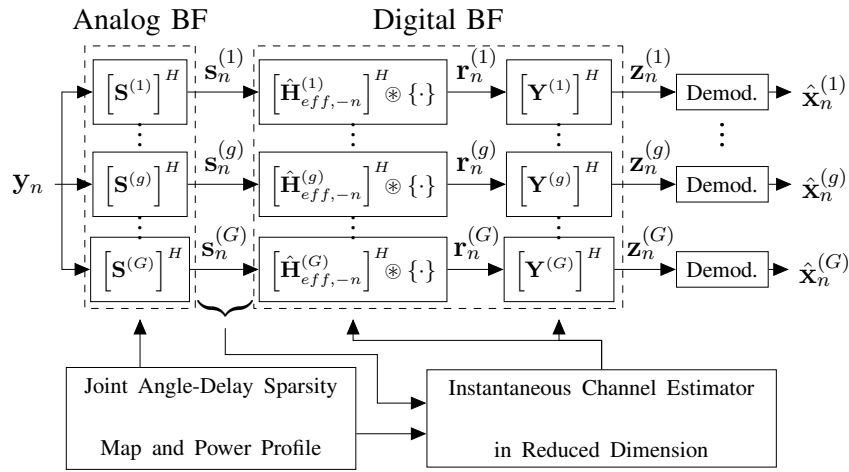
\begin{figure}[htbp]
\centering
\resizebox{0.7\textwidth}{!}{%
\begin{tikzpicture}[>=triangle 45]
[
xscale = 1, % to scale horizontally everything but the text
yscale = 1, % to scale vertically everything but the text
]
\def\xa{-5.2}
\def\xb{-5.1}
\def\xc{-4}
\def\xd{-2.8}
\def\xe{-1}
\def\xf{0.75}
\def\xg{1.9}
\def\xh{3.2}
\def\xi{4.2}
\def\xj{5.65}
\def\ya{2.25}
\def\yb{1.6}
\def\yc{1.1}
\def\yd{0.3}
\def\ye{-0.2}
\def\yf{-0.9}
\def\yg{-1.45}
\def\yh{-1.9}
\def\yhh{-2.1}
\def\yi{-3}
\def\xjadsm{-3}
\def\xicerd{2.5}
\node (n11) at (\xa,\yb)  {}; 
\node (n12) at (\xb,\yb) [coordinate] {};   
\node (n13) [block] at (\xc,\yb) {$\mbox{\footnotesize $ \left[ \mathbf{S}^{(1)} \right]^{H}$} \!$};  
\node (n14) [above] at (\xd,\yb) {$\mathbf{s}_n^{(1)}$}; 
\node (n15) [block]  at (\xe,\yb) {$\mbox{\footnotesize $\left[ \hat{\mathbf{H}}_{eff,-n}^{(1)} \right]^{H} \! \! \! \oast \lbrace \cdot \rbrace $}$};
\node (n16) [above] at (\xf,\yb) {$\mathbf{r}_n^{(1)}$};
\node (n17) at (\xg,\yb) [block] {$\mbox{\footnotesize $\left[ {\mathbf{Y}}^{(1)} \right]^{H}$}$}; 
\node (n18) at (\xh,\yb) [above] {$\mathbf{z}_n^{(1)}$}; 
\node (n19) [block] at (\xi,\yb) {\footnotesize{Demod.}}; 
\node (n110) at (\xj,\yb) {$\hat{\mathbf{x}}_n^{(1)}$}; 

\node (n23) at (\xc,\yc) {$\vdots$};  
\node (n25) at (\xe,\yc) {$\vdots$};  
\node (n27) at (\xg,\yc) {$\vdots$}; 
\node (n27) at (\xi,\yc) {$\vdots$}; 

% —————————— row 1
\node (n31) at (\xa,\yd) [left] {$\mathbf{y}_n$}; 
\node (n32) at (\xb,\yd) [coordinate] {}; 
\node (n33) at (\xc,\yd) [block] {$\mbox{\footnotesize $\left[ \mathbf{S}^{(g)} \right]^{H}$} \!$};  
\node (n34) at (\xd,\yd) [above]{$\mathbf{s}_n^{(g)}$}; 
\node (n35) at (\xe,\yd) [block] {$\mbox{\footnotesize $\left[ \hat{\mathbf{H}}_{eff,-n}^{(g)} \right]^{H} \! \! \! \oast \lbrace \cdot \rbrace $}$}; 
\node (n36) at (\xf,\yd) [above] {$\mathbf{r}_n^{(g)}$}; 
\node (n37) at (\xg,\yd) [block] {$\mbox{\footnotesize $\left[ {\mathbf{Y}}^{(g)} \right]^{H}$}$}; 
\node (n38) at (\xh,\yd) [above] {$\mathbf{z}_n^{(g)}$}; 
\node (n39) at (\xi,\yd) [block] {\footnotesize{Demod.}}; 
\node (n310) at (\xj,\yd) {$\hat{\mathbf{x}}_n^{(g)}$};

\node (n43) at (\xc,\ye) {$\vdots$};  
\node (n45) at (\xe,\ye) {$\vdots$}; 
\node (n47) at (\xg,\ye) {$\vdots$};
\node (n27) at (\xi,\yc) {$\vdots$}; 
% —————————— row 1
\node (n51) at (\xa,\yf) [coordinate] {}; 
\node (n52) at (\xb,\yf) [coordinate] {};  
\node (n53) at (\xc,\yf) [block] {$\mbox{\footnotesize $\left[ \mathbf{S}^{(G)} \right]^{H}$} \!$};  
\node (n54) at (\xd,\yf) [above] {$\mathbf{s}_n^{(G)}$}; 
\node (n55) at (\xe,\yf) [block] {$\mbox{\footnotesize $\left[ \hat{\mathbf{H}}_{eff,-n}^{(G)} \right]^{H} \! \! \! \oast \lbrace \cdot \rbrace $}$}; 
\node (n56) at (\xf,\yf) [above] {$\mathbf{r}_n^{(G)}$};
\node (n57) at (\xg,\yf) [block] {$\mbox{\footnotesize $\left[ {\mathbf{Y}}^{(G)} \right]^{H}$}$}; 
\node (n58) at (\xh,\yf) [above] {$\mathbf{z}_n^{(G)}$};
\node (n59) at (\xi,\yf) [block] {\footnotesize{Demod.}};
\node (n510) at (\xj,\yf) {$\hat{\mathbf{x}}_n^{(G)}$}; 
\node (d112) at (\xc,\ya) {};
\node (d123) at (\xd,\ya)  {}; 
\node (d134) at (\xe,\ya)  {}; 
\node (d145) at (\xg,\ya) {};
\node (d212) at (\xc,\yg) {}; 
\node (d223) at (\xd,\yg) {$\underbrace{\hspace{0.2cm}}$}; 
\node (d234) at (\xe,\yg) {};
\node (d245) at (\xg,\yg) {};
\node (b1) [l2block] at (\xjadsm,\yi) {\footnotesize{Joint Angle-Delay Sparsity Map and Power Profile}};
\node (b2) [l3block] at (\xicerd,\yi) {\footnotesize{Instantaneous Channel Estimator  in Reduced Dimension}};
\node (d11) [ left=0.63cm of d112, coordinate] {};
\node (d12) [right=0.63cm of d112, coordinate] {};
\node (d13) [ left=1.25cm of d134, coordinate] {};
\node (d14) [right=0.75cm of d145, coordinate] {};
\node (d21) [ left=0.63cm of d212, coordinate] {};
\node (d22) [right=0.63cm of d212, coordinate] {};
\node (d23) [ left=1.25cm of d234, coordinate] {};
\node (d24) [right=0.75cm of d245, coordinate] {};
\draw[-] (n31) |- (n32);
\draw[->] (n32) -- (n12) -- (n13);
\draw[->] (n32) -- (n33);
\draw[->] (n32) -- (n52) -- (n53);
\draw[->] (n13) -- (n15);
\draw[->] (n33) -- (n35);
\draw[->] (n53) -- (n55);
\draw[->] (n15) -- (n17);
\draw[->] (n35) -- (n37);
\draw[->] (n55) -- (n57);
\draw[->] (n17) -- (n19);
\draw[->] (n37) -- (n39);
\draw[->] (n57) -- (n59);
\draw[->] (n19) -- (n110);
\draw[->] (n39) -- (n310);
\draw[->] (n59) -- (n510);
\draw[->] (b1.north -| d212) -- (d212);
\draw[->] (b2.north) -- (\xicerd,\yh) -| (d234);
\draw[->] (b2.north) -- (\xicerd,\yh) -| (d245);
\draw[->] (b1.east |- b2.185) -- (b2.185);
\draw[->] (d223) -- (\xd,\yhh) -| ($(b1.north east)!0.3!(b2.north west)$ ) |- (b2.175);
\draw[dashed] (d12) -- (d22) -- (d21) -- (d11) -- (d12) node [above, pos = 0.5] {Analog BF};
\draw[dashed] (d14) -- (d24) -- (d23) -- (d13) -- (d14) node [above, pos = 0.5] {Digital BF};
\end{tikzpicture}
}
\caption{Equivalent receiver processing}
\label{Rx}
\end{figure}

Baseband digital equivalent receiver processing is shown in Fig. \ref{Rx}. The block \emph{Joint Angle-Delay Sparsity Map and Power Profile} is studied in \cite{Kalayci19}. It estimates center angle $\hat\mu_{\phi,l}^{(g)}$, angular spread $\hat\sigma_{\phi,l}^{(g)}$, and power profile $\hat{\rho}_{\phi,l}^{(g)}(\phi)$ for all MPCs of all users. With this information, CCM estimate $\hat{\mathbf{R}}_l^{(g)}$ can be calculated parametrically in a way similar to (\ref{R_gen_int}).

Analog beamformer is a bank of filters each of which projects $N$-dimensional observation into group-specific $D_g$-dimensional subspaces where $D_g$ is the number of RF chains allocated for the group $g$.  Analog beamformer selects intended group signals, rejects interference due to other groups (i.e., inter-group interference, IGI) and separates MPCs in angular domain. The analog beamformer matrix $\mathbf{S}^{(g)}$ is of size $N \times D_g$ where $D_g < N$. The output of analog beamformer stage for a chosen intended group $\tilde{g}$ is
\begin{equation} \label{gen_reduced}
\mathbf{s}_n^{(\tilde{g})}\triangleq\left[\mathbf{S}^{(\tilde{g})}\right]^{H}\mathbf{y}_n.
\end{equation} 

At this stage, effective channel and effective covariance matrices are defined as
 \begin{align}
 \mathbf{H}_{eff,l}^{(g)} & \triangleq \left[\mathbf{S}^{(\tilde{g})}\right]^{H} \mathbf{H}_l^{(g)},
 \\
 \mathbf{R}_{eff,l}^{(g)} & \triangleq \left[\mathbf{S}^{(\tilde{g})}\right]^{H} \mathbf{R}_{l}^{(g)} \mathbf{S}^{(\tilde{g})} , \label{R_eff_l_g}
 \\
  \mathbf{R}_{eff,\eta}^{(g)} & \triangleq \left[\mathbf{S}^{(\tilde{g})}\right]^{H} \mathbf{R}_{\eta}^{(g)} \mathbf{S}^{(\tilde{g})}, \label{R_eff_eta_g}
 \end{align}
for $g=1,\dots, G$. $\hat{\mathbf{H}}_{eff,l}^{(g)}$, $\hat{\mathbf{R}}_{eff,l}^{(g)}$ and $\hat{\mathbf{R}}_{eff,\eta}^{(g)}$  are their estimates. $\hat{\mathbf{H}}_{eff,l}^{(g)}$ is directly estimated from $\mathbf{s}_n^{(g)}$ by \emph{Instantaneous Channel Estimator in Reduced Dimension}. $\hat{\mathbf{R}}_{eff,l}^{(g)}$ and $\hat{\mathbf{R}}_{eff,\eta}^{(g)}$ are obtained by \eqref{R_eta}, \eqref{R_eff_l_g} and \eqref{R_eff_eta_g} using $\hat{\mathbf{R}}_{l}^{(g)}$, parametrically calculated  as \eqref{R_gen_int} with estimates $\hat{\mu}_{\phi,l}^{(g)}$ and $\hat{\sigma}_{\phi,l}^{(g)}$.
 
The output of the digital beamformer (constructed as channel matched filter, CMF) for the intended group is \begin{equation} \label{r_n}
 \mathbf{r}_n^{(\tilde{g})} \triangleq \sum_{l=0}^{L-1}\left[\hat{\mathbf{H}}_{eff,l}^{(\tilde{g})}\right]^{H}\mathbf{s}_{n+l}^{(\tilde{g})} ,
\end{equation} where $\hat{\mathbf{H}}_{eff,l}^{(\tilde{g})}$ is the estimated effective channel provided by \emph{Instantaneous Channel Estimator in Reduced Dimension}. As it is seen, (\ref{r_n}) is a time convolution and $\oast$ in Figure \ref{Rx} indicates this operation. Additionally, for intra-beam multiplexing purposes, a spatial zero forcing (SZF) filter $\mathbf{Y}^{(g)}$ can be applied when there are multiple users in the group. It outputs
\begin{equation}
\label{z_n}
\mathbf{z}_n^{(\tilde{g})} \triangleq \left[ \mathbf{Y}^{(\tilde{g})} \right]^{H} \mathbf{r}_n^{(\tilde{g})}
\end{equation}
where 
\begin{equation}
\label{Y_g}
\mathbf{Y}^{(\tilde{g})} \triangleq  \left\lbrace \begin{matrix} \mathbf{R}_0^{\hat{h} \hat{h}} \left( \left[ \mathbf{R}_0^{\hat{h} \hat{h}} \right]^{H} \mathbf{R}_0^{\hat{h} \hat{h}} \right)^{-1} , & \text{if} \, K_{\tilde{g}} > 1 \\
1 , & \text{if} \, K_{\tilde{g}} = 1 \end{matrix} \right.
\end{equation}
\begin{equation}
\mathbf{R}_l^{\hat{h} \hat{h}} \triangleq \sum_{l_1=0}^{L-1} \left[ \hat{ \mathbf{H}}_{eff,l_1}^{(\tilde{g})} \right]^{H} \hat{ \mathbf{H}}_{eff,l+l_1}^{(\tilde{g})}
\end{equation}

The output of the digital beamformer for the $\tilde{k}$\textsuperscript{th} user can be categorized as
\begin{equation} \label{zseparation}
\begin{aligned}
z_n^{(\tilde{g}_{\tilde{k}})} =& \underbrace{\mathbf{e}_{\tilde{k}}^{H} \left[ \mathbf{Y}^{(\tilde{g})}\right]^{H} \mathbf{R}_0^{\hat{h}\hat{h}} \mathbf{e}_{\tilde{k}} \, x_{n}^{(\tilde{g}_{\tilde{k}})}}_{\text{Intended Signal (S)}} + \underbrace{\mathbf{e}_{\tilde{k}}^{H} \left[ \mathbf{Y}^{(\tilde{g})}\right]^{H} \left(\mathbf{R}_0^{\hat{h} h} - \mathbf{R}_0^{\hat{h} \hat{h}} \right) \mathbf{e}_{\tilde{k}} \, x_{n}^{(\tilde{g}_{\tilde{k}})}}_{\substack{\text{Self-Interference due to} \\ \text{Ch. Est. Error (SICEE)}}} 
\\
 +& \underbrace{\sum_{\substack{l=-(L-1) \\ l\neq0}}^{L-1} \mathbf{e}_{\tilde{k}}^{H} \left[ \mathbf{Y}^{(\tilde{g})}\right]^{H} \mathbf{R}_l^{\hat{h}h} \mathbf{e}_{\tilde{k}} \, x_{n-l}^{(\tilde{g}_{\tilde{k}})}}_{\text{Inter-Symbol Int. (ISI)}} 
 + \underbrace{\sum_{l=-(L-1)}^{L-1} \sum_{\substack{k'=1 \\ k' \neq \tilde{k}}}^{K} \mathbf{e}_{\tilde{k}}^{H} \left[ \mathbf{Y}^{(\tilde{g})}\right]^{H} \mathbf{R}_l^{\hat{h}h} \mathbf{e}_{k'} \, x_{n-l}^{(\tilde{g}_{k'})}}_{\text{Multi-User Int. (MUI)}} \\ 
 +& \underbrace{\sum_{l=0}^{L-1} \mathbf{e}_{\tilde{k}}^{H} \left[ \mathbf{Y}^{(\tilde{g})}\right]^{H} \left[\hat{\mathbf{H}}_{eff,l}^{(\tilde{g})}\right]^{H} \left[\mathbf{S}^{(\tilde{g})}\right]^{H} \bm{\xi}_{n+l}}_{\text{Inter-Group Int. (IGI)}} + \underbrace{\sum_{l=0}^{L-1} \mathbf{e}_{\tilde{k}}^{H} \left[ \mathbf{Y}^{(\tilde{g})}\right]^{H} \left[\hat{\mathbf{H}}_{eff,l}^{(\tilde{g})}\right]^{H} \left[\mathbf{S}^{(\tilde{g})}\right]^{H} \mathbf{n}_{n+l}}_{\text{Noise}},
\end{aligned}
\end{equation}
where $\mathbf{e}_{k}$ is the $k$\textsuperscript{th} column of $\mathbf{I}_{K_{\tilde{g}}}$ for $k=1,...,K_{\tilde{g}}$ and
\begin{flalign}
\mathbf{R}_l^{\hat{h} h} & \triangleq \sum_{l_1=0}^{L-1} \left[ \hat{ \mathbf{H}}_{eff,l_1}^{(\tilde{g})} \right]^{H} \mathbf{H}_{eff,l+l_1}^{(\tilde{g})} 
\\
\bm{\xi}_{n}^{(\tilde{g})} & \triangleq \sum_{\substack{g'=1 \\ g' \neq \tilde{g}}}^{G} \sum_{l=0}^{L-1} \mathbf{H}_l^{(g')}\mathbf{x}_{n-l}^{(g')}.
\end{flalign}

Then, the output SINR for $\tilde{k}$\textsuperscript{th} user of $\tilde{g}$\textsuperscript{th} group is
\begin{equation} \label{sinr}
\text{SINR}^{(\tilde{g}_{\tilde{k}})}=\frac{E[\,|\, t_n^{(\tilde{g}_{\tilde{k}})} \, |^2 \,]}{E[\,|\, z_n^{(\tilde{g}_{\tilde{k}})} - t_n^{(\tilde{g}_{\tilde{k}})} \,|^2\,]}
\end{equation}
where $t_n^{(\tilde{g}_{\tilde{k}})}$ is the \emph{Intended Signal} term in \eqref{zseparation}. Detailed explanation about receiver processing can be found in \cite{kurt19, kurtmsc}.
 
 \subsubsection{Nearly Optimal Generalized Eigen-Beamformer} \label{secGEB}
\begin{figure}[htbp]
\centerline{\includegraphics[width=0.7\textwidth]{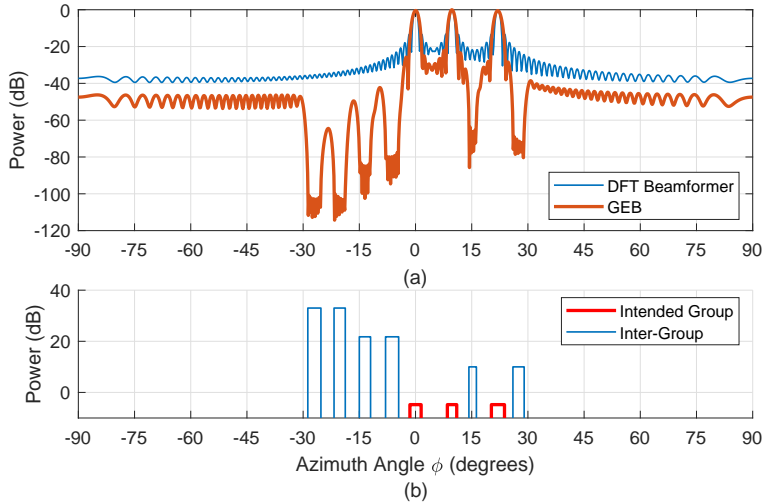}}
\caption{Comparison of GEB and DFT beamformers. (a) Power patterns. (b) Placement in the angular domain.}
\label{2Patterns}
\end{figure}

 In this method, CCMs related to the $l$\textsuperscript{th} delay are used to construct the beamformer related to the $m$\textsuperscript{th} MPC $\mathbf{S}_m^{(g)}$, where the relation between $m$ and $l$ is constituted by the function $\mathcal{L}^{(g)}(\cdot)$ as $l=\mathcal{L}^{(g)}(m)$. Generalized eigenvectors $\mathbf{v}_p$ of the matrix pair $(\hat{\mathbf{R}}_l^{(g)},\hat{\mathbf{R}}_y)$ which obey 
\begin{equation}
\hat{\mathbf{R}}_l^{(g)} \mathbf{v}_p = \lambda_p  \hat{\mathbf{R}}_y \mathbf{v}_p, \, \text{ for } \, p=1,\dots,N
\end{equation} are found where $\lambda_p$ is the $p$\textsuperscript{th} most dominant generalized eigenvalue which corresponds to $\mathbf{v}_p$. In addition, $\hat{\mathbf{R}}_l^{(g)}$ and $\hat{\mathbf{R}}_y$ are estimates of $\mathbf{R}_l^{(g)}$ and $\mathbf{R}_y$ which are defined in (\ref{R_gen_int}) and (\ref{Ry}). $\mathbf{S}_m^{(g)}$ is constructed with the most dominant $d_m^{(g)}$ generalized eigenvectors by horizontally concatenating them. Its beam pattern $\left| | \left[ \mathbf{S}^{(g)} \right]^{H} \mathbf{u}(\phi) | \right|^2$ is compared with that of the conventional DFT beamformer \cite{kurt19} in Fig. \ref{2Patterns}, where scenario in Table \ref{tb_angular_sec2} is employed for group 1 and $d_m^{(1)}=1$ for $m=1,2,3$ ($D_1=3$). 

Optimality of GEB is discussed in \cite{kurt19,kurtmsc}. As it is seen in Fig. \ref{2Patterns}, GEB exhibits a remarkable interference-cancelling ability. Also, it is notable that nulls are multiple and broad, which is rare in literature. Nulls having a nonzero width is indispensable in statistical beamforming with slow-time update, due to angular spread phenomenon. Another advantage is that each peak is coming from merely one column of $\mathbf{S}^{(g)}$ while other two have nulls there, which plays a significant role in ISI mitigation \cite{kurtmsc}.

Since angular spreads are small, it is sufficient to use only the most dominant generalized eigenvector as $\mathbf{S}_m^{(g)}=\mathbf{v}_1$. In addition, we can use rank-one approximation for $\hat{\mathbf{R}}_l^{(g)}$. Then, as a suboptimal solution, $\mathbf{S}_m^{(g)}$ can be constructed as  
\begin{equation}\label{Ryinvu}
\mathbf{S}_m^{(g)}=\hat{\mathbf{R}}_y^{-1}\,\hat{\mathbf{w}}_{1,l}^{(g)}
\end{equation}
where $l=\mathcal{L}^{(g)}(m)$, and $\hat{\mathbf{w}}_{1,l}^{(g)}$ is the estimation to the first eigenvector of $\mathbf{R}_l^{(g)}$ as defined in (\ref{gevecRlg}).
Then, analog beamformer matrix $\mathbf{S}^{(g)}$ is constructed from $\mathbf{S}_m^{(g)}$ for $m=1,\dots,M^{(g)}$ as \begin{equation} \label{abf_conf}
 \mathbf{S}^{(g)} = \left[ \begin{matrix} & \mathbf{S}_1^{(g)} & \mathbf{S}_2^{(g)} & \cdots & \mathbf{S}_{M^{(g)}}^{(g)} & \end{matrix} \right].
\end{equation} The matrix $\mathbf{S}^{(g)}$ is of size $N \times D_g$ where $D_g=\sum_{m=1}^{M^{(g)}} d_m^{(g)}$ is the number of RF-chains allocated for the group $g$. Note that the estimated CCMs of each MPC, namely $\hat{\mathbf{R}}_l^{(g)}$, are constructed parametrically as in (\ref{R_gen_int}). Moreover, $\hat{\mathbf{R}}_y$ is found by replacing $\mathbf{R}_l^{(g)}$ with $\hat{\mathbf{R}}_l^{(g)}$ in (\ref{Ry}).

\section{Proposed Adaptive Beamformer Construction Methods}
\subsection{GEB with Recursively Filtered Channel Covariance Matrices} \label{sec_geb_rec_filt}
In the previous work \cite{kurt19}, it is shown that the GEB is a powerful method, but vulnerable against the bias in the angular estimations. This method aims to exploit the advantages of the GEB in the presence of erroneous angular estimations by benefiting from the correlation of AoA in slow-time. It is proposed to construct GEB with recursively filtered CCMs, instead of the last estimation. The filtered covariance matrices are obtained as
\begin{equation}
\hat{\mathbf{R}}_l^{(g),f}[n] \triangleq \beta \, \hat{\mathbf{R}}_l^{(g),f}[n-1] + \left( 1-\beta \right) \hat{\mathbf{R}}_l^{(g)}[n]
\end{equation}
and 
\begin{equation}
\hat{\mathbf{R}}_y^{f}[n] \triangleq \beta \hat{\mathbf{R}}_y^{f}[n-1] + \left( 1-\beta \right) \hat{\mathbf{R}}_y [n] \text{ for } n=1,2,3,...
\end{equation}
where $0 \le \beta<1$. Then, they are used in GEB construction, using generalized eigendecomposition of the filtered matrix pair $(\hat{\mathbf{R}}_l^{(g),f}[n],\hat{\mathbf{R}}_y^{f}[n])$, instead of $(\hat{\mathbf{R}}_l^{(g)}[n],\hat{\mathbf{R}}_y[n])$. Note that $\hat{\mathbf{R}}_y^{f}[n]$ can also be found using $\hat{\mathbf{R}}_l^{(g),f}[n]$, similar to (\ref{Ry}).

\subsection{Wiener Filter Type Analog Beamformer}
This beamformer has the structure of the Wiener filter. However, operands are recursively filtered in time. Moreover, the correlation matrix inverse is updated via Woodbury matrix identity. The derivation benefits from \emph{Hadamard-factorization} technique in \eqref{eqhadamardfac} and \emph{angular patching} idea in \eqref{eqangularpatching} applied on CCMs, and mainly the resultant structure in \eqref{generalpatching}, explained in Appendix. To lower the computational complexity, some low-rank approximations and quantization on patch power levels are performed. Consider a modification on \eqref{generalpatching} as
\begin{equation}
\hat{\mathbf{R}}_l^{(g)} = \left(\mathbf{Q} \hat{\mathbf{P}}_l^{(g)} \mathbf{Q}^{H} \right) \odot \mathbf{D}_r,
\end{equation}
where $\mathbf{D}_r$ is the r-rank approximation to $\mathbf{D}$ using the expression in (\ref{eigenvectorsofD1}), and $\hat{\mathbf{P}}_l^{(g)}$ is obtained as explained in Appendix. Recursive filtering can be equivalently applied on estimated power levels $\hat{\mathbf{P}}_l^{(g)}$ instead of $\hat{\mathbf{R}}_l^{(g)}$ as 
\begin{equation}
\hat{\mathbf{P}}_l^{(g),f}=\beta \hat{\mathbf{P}}_l^{(g),f} + (1-\beta) \hat{\mathbf{P}}_l^{(g)}.
\end{equation}
Then, patch power levels on the diagonal matrix $\hat{\mathbf{P}}_l^{(g),f}$ are quantized such that n\textsuperscript{th} patch power level becomes
\begin{equation} \label{quantization2}
\left(\hat{\mathbf{P}}_l^{(g),q}\right)_{n,n} = c_{n} \frac{h_l^{(g)}}{N_q} \, \text{ for } \, n=1,2,\dots,N
\end{equation}
where $\frac{h_l^{(g)}}{N_q}$ is the width of quantization levels, $c_{n}$ is the chosen level which is found as
\begin{equation} \label{quantization3}
c_{n} \triangleq \underset{\tilde{c}}{\operatorname{argmin}} \left|\left(\hat{\mathbf{P}}_l^{(g),f}\right)_{n,n} - \tilde{c}\frac{h_l^{(g)}}{N_q} \right| \, \text{ for } \, \tilde{c} \in \mathbb{Z},
\end{equation}
and $h_l^{(g)} \triangleq \text{tr}\left(\mathbf{R}_l^{(g)} \right)\left(\left\lceil \sigma_{\theta,l}^{(g)} / \frac{2\pi}{N} \right\rceil \right)^{-1}$ is the expected single patch power level, and $N_q$ is quantization parameter. Note that the quantizers are MPC-specific because of $h_l^{(g)}$. Then, quantized observation patch power matrix $\hat{\mathbf{P}}_y^{q}$ can be obtained as 
\begin{equation}\label{PyfromPlg}
 \hat{\mathbf{P}}_y^{q} \triangleq \sum_{g=1}^G K_{g}E_s^{(g)} \sum_{l=0}^{L-1} \hat{\mathbf{P}}_{l}^{(g),q} + N_0 \mathbf{I}_N,
\end{equation} 
in accordance with (\ref{Ry}). Then, the quantized observation CCM can be expressed as
\begin{equation}\label{generalpatchingquantized}
\hat{\mathbf{R}}_y^{q} = \left(\mathbf{Q} \hat{\mathbf{P}}_y^{q} \mathbf{Q}^{H} \right) \odot \mathbf{D}_r.
\end{equation}

Expressing the above equation in terms of vectors and using the identity $\left(\mathbf{a}\mathbf{a}^{H}\right) \odot \left(\mathbf{b}\mathbf{b}^{H}\right) = \left(\mathbf{a} \odot \mathbf{b}\right) \left(\mathbf{a} \odot \mathbf{b}\right)^{H}$, an equivalent expression is obtained as
\begin{equation} \label{oversampled}
\hat{\mathbf{R}}_y^{q}=\mathbf{V} \bar{\mathbf{P}}_y^{q} \mathbf{V}^{H},
\end{equation}
where
\begin{equation}
\mathbf{V} \triangleq \left[
\begin{matrix} &
\mathbf{V}_1  &
\mathbf{V}_2  &
\cdots  &
\mathbf{V}_N &
\end{matrix}
\right],
\end{equation}
\begin{equation} \label{V_k}
\mathbf{V}_k \triangleq \left[ \begin{matrix} &
\mathbf{q}_k \odot \mathbf{d}_1  &
\mathbf{q}_k \odot \mathbf{d}_2  &
\cdots  &
\mathbf{q}_k \odot \mathbf{d}_r  &
\end{matrix}
\right],
\end{equation}
\begin{equation} \label{enlargep}
\bar{\mathbf{P}}_y^{q} \triangleq \hat{\mathbf{P}}_y^{q} \otimes \mathbf{I}_r ,
\end{equation}
and $\otimes$ is the Kronecker product operator. $\mathbf{q}_k$ is the k\textsuperscript{th} column of $\mathbf{Q}$ and $\mathbf{d}_n$ are weighted eigenvectors of $\mathbf{D}$ similar to \eqref{eigenvectorsofD1}. Note that the matrix $\mathbf{V}$ is of size $N \times Nr$, and the matrix $\bar{\mathbf{P}}_y^{q}$ is of size $Nr \times Nr$.

After that, consider the series of matrices $\hat{\mathbf{R}}_y^{q}[n]$ and correspondent $\hat{\mathbf{P}}_y^{q}[n]$, where $n$ is slow-time index. Let the update matrices be defined as $\Delta\hat{\mathbf{R}}_y^{q}[n] \triangleq \hat{\mathbf{R}}_y^{q}[n] - \hat{\mathbf{R}}_y^{q}[n-1]$ and
\begin{equation}
\Delta\hat{\mathbf{P}}_y^{q}[n] \triangleq \hat{\mathbf{P}}_y^{q}[n] - \hat{\mathbf{P}}_y^{q}[n-1].
\end{equation}
The matrix pair $\Delta\hat{\mathbf{R}}_y^{q}[n]$ and $\Delta\hat{\mathbf{P}}_y^{q}[n]$ fits into the structure in (\ref{oversampled}) as 
\begin{equation} \label{eqenlargedeltar}
\Delta\hat{\mathbf{R}}_y^{q}[n]=\mathbf{V} \Delta\bar{\mathbf{P}}_y^{q}[n] \mathbf{V}^{H},
\end{equation}
where $
\Delta\bar{\mathbf{P}}_y^{q}[n] \triangleq \Delta\hat{\mathbf{P}}_y^{q}[n] \otimes \mathbf{I}_r$. 

$\Delta \bar{\mathbf{P}}_y^{q}[n]$ involves the change in the patch powers. Then, only the powers of patches experiencing a movement are nonzero. This leads to an equivalent expression of (\ref{eqenlargedeltar}) as 
\begin{equation} \label{oversamplednz}
\Delta\hat{\mathbf{R}}_y^{q}[n]=\mathbf{V}_{nz} \Delta\bar{\mathbf{P}}_{y,nz}^{q}[n] \mathbf{V}_{nz}^{H},
\end{equation}
where $N_{\Delta p}$ is the number of nonzero patch powers, the matrix $\mathbf{V}_{nz}$ is of size $N \times N_{\Delta p}r$, and the matrix $\bar{\mathbf{P}}_{y,nz}^{q}[n]$ is of size $N_{\Delta p}r \times N_{\Delta p}r$. Then the inverse can be expressed as
\begin{equation}
\left(\hat{\mathbf{R}}_y^{q}[n]\right)^{-1} = \left( \hat{\mathbf{R}}_y^{q}[n-1] + \mathbf{V}_{nz} \Delta\bar{\mathbf{P}}_{y,nz}^{q}[n] \mathbf{V}_{nz}^{H} \right)^{-1}.
\end{equation}

If $( \, \hat{\mathbf{R}}_y^{q}[n-1] \, )^{-1}$ is known, $( \, \hat{\mathbf{R}}_y^{q}[n] \, )^{-1}$ can be found through Woodbury matrix identity as 
\begin{flalign} \label{woodbury} \small
\left(\hat{\mathbf{R}}_y^{q}[n]\right)^{ \hspace{-3pt}-1} \hspace{-5pt} = & \left(\hat{\mathbf{R}}_y^{q}[n-1]\right)^{ \hspace{-3pt}-1} \hspace{-5pt} \\
- & \left(\hat{\mathbf{R}}_y^{q}[n-1]\right)^{ \hspace{-3pt}-1} \hspace{-4pt} \mathbf{V}_{nz} \left( \left( \Delta\bar{\mathbf{P}}_{y,nz}^{q}[n] \right)^{\hspace{-2pt} -1} \hspace{-6pt} + \mathbf{V}_{nz}^{H} \left(\hat{\mathbf{R}}_y^{q}[n-1]\right)^{\hspace{-3pt}-1} \hspace{-5pt} \mathbf{V}_{nz} \right)^{\hspace{-4pt}-1} \hspace{-6pt} \mathbf{V}_{nz}^{H} \left(\hat{\mathbf{R}}_y^{q}[n-1]\right)^{\hspace{-3pt}-1} \normalsize \nonumber
\end{flalign}

The matrix inverse in the right hand side of (\ref{woodbury}) is on a matrix of size $N_{\Delta p} r \times N_{\Delta p} r$. The direct calculation of the matrix $( \, \hat{\mathbf{R}}_y^{q}[n] \, )^{-1}$ would require a matrix inversion operation of size $N \times N$.
$N$ and $N_{\Delta p} r$ will be compared in Section \ref{secresults}.

When used after a patching process, this method can update the inverse of a matrix series $\hat{\mathbf{R}}_y^{q}[n]$ with the input of $\Delta \mathbf{P}_y^{q}[n] $, promising low-complexity operation if the change rate of $\hat{\mathbf{R}}_y^{q}[n]$ is slow. 

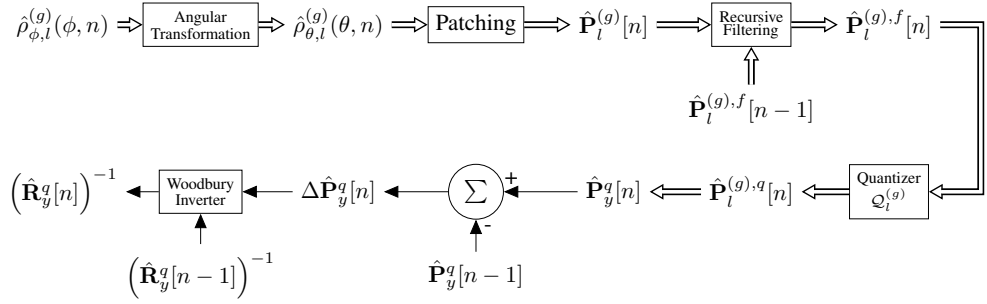
\begin{figure}[htbp]
\centering
\resizebox{0.8\textwidth}{!}{%
\begin{tikzpicture}[>=triangle 45]
[
xscale = 1, % to scale horizontally everything but the text
yscale = 1, % to scale vertically everything but the text
]
\small
\def\scscx{0.7}
\def\scscy{0.85}
\def\xaa{-3 * \scscx}
\def\xa{0 * \scscx}
\def\xb{3 * \scscx}
\def\xc{6 * \scscx}
\def\xd{9 * \scscx}
\def\xe{12 * \scscx}
\def\xf{15 * \scscx}
\def\xg{18 * \scscx}
\def\xh{20 * \scscx}

\def\ya{3 * \scscy}
\def\yaa{1.5 * \scscy}
\def\yb{0 * \scscy}
\def\ybb{-1.5 * \scscy}

\node (rhoph) [] at (\xa,\ya)  {$\hat{\rho}_{\phi,l}^{(g)}(\phi,n)$};

\node (angtra) [block] at (\xb,\ya)  {$\substack{\text{Angular} \\ \text{Transformation}}$};

\node (rhoth) [] at (\xc,\ya)  {$\hat{\rho}_{\theta,l}^{(g)}(\theta,n)$};

\node (patching) [block] at (\xd,\ya)  {Patching};

\node (Phat) [] at (\xe,\ya) {$\hat{\mathbf{P}}_{l}^{(g)}[n]$};

\node (Pfilt2) [] at (\xf,\yaa) {$\hat{\mathbf{P}}_{l}^{(g),f}[n-1]$};

\node (recfilt) [block] at (\xf,\ya)  {$\substack{\text{Recursive} \\ \text{Filtering}}$};

\node (Pfilt) [] at (\xg,\ya)  {$\hat{\mathbf{P}}_{l}^{(g),f}[n]$};

\node (quant) [block] at (\xg,\yb)  {$\substack{\text{Quantizer} \\ \mathcal{Q}_{l}^{(g)}}$};

\node (Pquant) [] at (\xf,\yb)  {$\hat{\mathbf{P}}_{l}^{(g),q}[n]$};

\node (Pyquant) [] at (\xe,\yb)  {$\hat{\mathbf{P}}_{y}^{q} [n]$};

\node (Pyquant2) [] at (\xd,\ybb)  {$\hat{\mathbf{P}}_{y}^{q} [n-1]$};

\node (differ) [summer] at (\xd,\yb) {$\sum$};

\node (dPyquant) [] at (\xc,\yb)  {$\Delta\hat{\mathbf{P}}_{y}^{q} [n]$};

\node (invert) [block] at (\xb,\yb) {$\substack{\text{Woodbury} \\ \text{Inverter}}$};

\node (rinv2) [] at (\xb,\ybb)  {$\left( \hat{\mathbf{R}}_{y}^{q} [n-1] \right)^{-1}$};

\node (rinv) [] at (\xa,\yb)  {$\left( \hat{\mathbf{R}}_{y}^{q} [n] \right)^{-1}$};
%\node (ccmcon) [block, text width=5em] at (\xj,\ya)  {$\substack{\text{CCM} \\ \text{Construction}}$};
%
%\node (Rquant) [] at (\xk,\ya) {$\hat{\mathbf{R}}_{l}^{(g),q}[n]$};

\draw[vecArrow] (rhoph) -- (angtra);
\draw[vecArrow] (angtra) -- (rhoth);
\draw[vecArrow] (rhoth) -- (patching);
\draw[vecArrow] (patching) -- (Phat);
\draw[vecArrow] (Phat) -- (recfilt);
\draw[vecArrow] (Pfilt2) -- (recfilt);
\draw[vecArrow] (recfilt) -- (Pfilt);
\draw[vecArrow] (Pfilt) -- (\xh,\ya) --(\xh,\yb) -- (quant);
\draw[vecArrow] (quant) -- (Pquant);
\draw[vecArrow] (Pquant) -- (Pyquant);
\draw[->] (Pyquant) -- (differ) node[pos=0.9, above] {+};
\draw[->] (Pyquant2) -- (differ) node[pos=0.75, right] {-};
\draw[->] (differ) -- (dPyquant);
\draw[->] (dPyquant) -- (invert);
\draw[->] (rinv2) -- (invert);
\draw[->] (invert) -- (rinv);
%\draw[->] (Pquant) -- (ccmcon);
%\draw[->] (ccmcon) |- (Rquant);
\end{tikzpicture}
}
\caption{Efficient analog beamformer construction: Adaptive calculation of the inverse term}
\label{figproposedadaptivewiener}
\end{figure}

It is proposed to construct the MPC-specific analog beamformer for the $n$\textsuperscript{th} slow-time instance $\mathbf{S}_m^{(g)}[n]$ as 
\begin{equation}
\mathbf{S}_m^{(g)}[n]=\left( \hat{\mathbf{R}}_y^{q}[n] \right)^{-1} \hat{\mathbf{w}}_{1,l}^{(g),f}[n],
\end{equation}
where $( \, \hat{\mathbf{R}}_y^{q}[n] \, )^{-1}$ is obtained using the aforementioned processing blocks as shown in Figure \ref{figproposedadaptivewiener}. On the other hand,
\begin{equation} \label{steering_vec_filtering}
\hat{\mathbf{w}}_{1,l}^{(g),f}[n]\triangleq \beta \hat{\mathbf{w}}_{1,l}^{(g),f}[n-1] + (1 - \beta) \hat{\mathbf{w}}_{1,l}^{(g)}[n]
\end{equation}
where $\hat{\mathbf{w}}_{1,l}^{(g)}[n]$ is the estimate for $\mathbf{w}_{1,l}^{(g)}[n]$ in (\ref{gevecRlg}). It is found by using $\hat{\mu}_{\theta,l}^{(g)}[n]$ and $\hat{\sigma}_{\theta,l}^{(g)}[n]$ instead of $\mu_{\theta,l}^{(g)}[n]$ and $\sigma_{\theta,l}^{(g)}[n]$. After constructing $\mathbf{S}_m^{(g)}[n]$ for $m=1,\dots,M^{(g)}$, $\mathbf{S}^{(g)}[n]$ is obtained as shown in (\ref{abf_conf}). If multiple RF chains are to be used for one MPC ($d_m^{(g)}>1$), $\hat{\mathbf{w}}_{n,l}^{(g),f}[n]$ for $n=1,...,d_m^{(g)}$ are used in construction similarly.

This construction process is depicted in Figure \ref{figproposedadaptivewiener}. This method has three parameters: $\beta$ from the recursive filtering of the patch power levels and the steering vector, $N_q$ from the quantizer, and $r$ from the reduced-rank expression $\mathbf{D}_r$ of the matrix $\mathbf{D}$ in Woodbury inverter.

\subsection{Whitening Filter Type Analog Beamformer}
This filter is a variant of the Wiener filter type analog beamformer. For this method, $\mathbf{S}_m^{(g)}[n]$ is constructed as 
\begin{equation}
\mathbf{S}_m^{(g)}[n]=\left( \hat{\mathbf{R}}_{\eta,l}^{(g),q}[n] \right)^{-1} \hat{\mathbf{w}}_{1,l}^{(g),f}[n],
\end{equation}
where $\hat{\mathbf{R}}_{\eta,l}^{(g),q}[n]\triangleq\hat{\mathbf{R}}_y^{q}[n] - E_s K_g \hat{\mathbf{R}}_l^{(g),q}[n]$ is used instead of $ \hat{\mathbf{R}}_y^{q}[n]$. This modification aims a performance increase by avoiding suppression of remaining eigenvectors of the intended signal by excluding them from $\hat{\mathbf{R}}_y^{q}[n]$, at a cost of extra computational complexity of obtaining $( \, \hat{\mathbf{R}}_{\eta,l}^{(g),q}[n] \, )^{-1}$ from $( \, \hat{\mathbf{R}}_y^{q}[n] \, )^{-1}$ through Woodbury inverter.

\subsubsection*{Complexity Analysis}
The size of matrix inversions is considered to be dominant and taken as the complexity measure for proposed processing schemes. It is $N_{\Delta p} r$ for Wiener filter type, and $( \, N_{\Delta p} + N_{p,l}^{(g)} \, ) r$ for whitening filter type analog beamformer construction, considering the conversion from $(\, \hat{\mathbf{R}}_y^{q}[n] \, )^{-1}$ to $( \, \hat{\mathbf{R}}_{\eta,l}^{(g),q}[n] \, )^{-1}$ where $N_{p,l}^{(g)}$ is the number of patches with nonzero powers related with $\hat{\mathbf{R}}_l^{(g),q}[n]$. Note that $N$ many power levels are processed until the inversion step, and quantized values of $\hat{\sigma}_{\theta,l}^{(g)}$ can be used for $\hat{\mathbf{w}}_{1,l}^{(g)}[n]$ in \eqref{gevecRlg} for \eqref{steering_vec_filtering}. 

For GEB, it is taken as $N$ due to the matrix inversion of size $N$ applied in algorithms for generalized eigendecomposition.

\section{Analytic Performance Measures}

\subsection{Analytical Expressions for Signal Powers at the Output of CMF}
Analytical expressions for signal powers can be provided for the output of the CMF (output of digital beamformer when SZF is not applied, or $K_{\tilde{g}}=1$). It is due to the fact that channels are assumed to be circularly symmetric complex normal random vectors. Therefore, the output is only a sum of products of normally distributed random variables.  Defining $\mathbf{R}_{eff,sum}^{(g)} \triangleq \sum_{l=0}^{L-1} \mathbf{R}_{eff,l}^{(g)} $, analytical expressions for $P_S^{(\tilde{g}),\text{CMF}}\triangleq E[\,|\, t_n^{(\tilde{g}_k)} \, |^2 \,]$ and $P_{IN}^{(\tilde{g}),\text{CMF}} \triangleq E[\,|\, z_n^{(\tilde{g}_k)} - t_n^{(\tilde{g}_k)} \,|^2\,]$ are given in (\ref{p_s_an}) and (\ref{p_in_an}), respectively; for the case that instantaneous channel is perfectly known ($\hat{\mathbf{H}}_{eff,l}^{(g)} = \mathbf{H}_{eff,l}^{(g)}$ but $\hat{\mathbf{R}}_{l}^{(g)} \neq \mathbf{R}_{l}^{(g)}$). They are used to calculate SINR as shown in \eqref{sinr}. The full derivation can be found in \cite{kurtmsc}.
\begin{align}
P_S^{(\tilde{g}),\text{CMF}} = & E_s^{(\tilde{g})} \left[ \text{tr}\left( \mathbf{R}_{eff,sum}^{(\tilde{g})} \right)^2
+ \sum_{l=0}^{L-1} \text{tr} \left( \mathbf{R}_{eff,l}^{(\tilde{g})} \mathbf{R}_{eff,l}^{(\tilde{g})} \right) \right] \label{p_s_an}
\\
P_{IN}^{(\tilde{g}),\text{CMF}} = &
\sum_{g' = 1}^G
E_s^{(g')}
\left[ \text{tr}\left( \mathbf{R}_{eff,sum}^{(\tilde{g})} \right)^2 + K_{g'} \text{tr}\left( \mathbf{R}_{eff,sum}^{(\tilde{g})} \mathbf{R}_{eff,sum}^{(g')} \right) \right]  \nonumber
\\
+ & N_0 
\text{tr}\left(\mathbf{R}_{eff,sum}^{(\tilde{g})} \left[\mathbf{S}^{(\tilde{g})}\right]^{H} \mathbf{S}^{(\tilde{g})} \right) - P_S^{(\tilde{g}),CMF}
\label{p_in_an}
\end{align}

\subsection{Channel Estimation}

Instantaneous channel estimator in reduced domain, shown in Figure \ref{Rx}, estimates effective channel matrix $\mathbf{H}_{eff,l}^{(\tilde{g})}$, hence effective channel vectors $\mathbf{h}_{eff,l}^{(\tilde{g}_k)}$. To do this, users in the same group is subjected to a training phase where they are transmitting a predetermined training sequence consisting of $T$ symbols. This training period does not need to be synchronized among groups and other groups might be in data mode, thanks to interference suppression of analog beamformer. 

It will be beneficial to express the variables to estimate as a single vector. Therefore, a new group channel vector notation is defined as \begin{equation} \label{h_eff_bar1}
\bar{\mathbf{h}}_{eff}^{(g)} \triangleq \left[\begin{matrix}\left[ \bar{\mathbf{h}}_{eff}^{(g_1)} \right]^{H} & \left[ \bar{\mathbf{h}}_{eff}^{(g_2)} \right]^{H} & \hdots & \left[ \bar{\mathbf{h}}_{eff}^{(g_{K_g})} \right]^{H} \end{matrix} \right]^{H} ,
\end{equation} where \begin{equation} \label{h_eff_bar2}
 \bar{\mathbf{h}}_{eff}^{(g_k)} \triangleq \left[\begin{matrix} \left[ \mathbf{h}_{eff,0}^{(g_k)} \right]^{H} & \left[ \mathbf{h}_{eff,1}^{(g_k)} \right]^{H} & \hdots & \left[ \mathbf{h}_{eff,L-1}^{(g_k)} \right]^{H} \end{matrix}  \right]^{H} .
\end{equation} 

Noting $\mathbf{h}_{eff,l}^{(g_k)} = \left[\mathbf{S}^{(\tilde{g})}\right]^{H} \mathbf{h}_{l}^{(g_k)}$, using $\bar{\mathbf{h}}_{eff}^{(g)}$ and obeying general system definitions given in (\ref{gen_full2}) and (\ref{gen_reduced}), the output of analog beamformer $\bar{\mathbf{s}}^{(\tilde{g})} \triangleq \left[ \left[\mathbf{s}_0^{(\tilde{g})}\right]^{H} \, ... \, \left[\mathbf{s}_{T-1}^{(\tilde{g})}\right]^{H} \right]^{H}$ throughout one training period of $T$ (i.e., for $n$ from $0$ to $T-1$) is expressed as 
\begin{equation}
 \bar{\mathbf{s}}^{(\tilde{g})} = \sum_{g=1}^G \left( \mathbf{X}^{(g)} \otimes \mathbf{I}_{D_{\tilde{g}}} \right) \bar{\mathbf{h}}_{eff}^{(g)} + \left( \mathbf{I}_T \otimes \left[ \mathbf{S}^{(\tilde{g})} \right]^{H} \right) \bar{\mathbf{n}} ,
\label{ch_est1} 
\end{equation} 
where $\otimes$ is the Kronecker product operator. The reduced-dimension observation vector $\bar{\mathbf{s}}^{(\tilde{g})}$ of size $T D_{\tilde{g}}$ and the noise vector $\bar{\mathbf{n}}$ of size $T N$ are vertically concatenated forms of reduced-dimension observation vector $\mathbf{s}_n^{(\tilde{g})}$ and full-dimension noise term $\mathbf{n}_n$, respectively, for $n$ from $0$ to $T-1$. The matrix $\mathbf{X}^{(g)}$ involves tranmitted symbols $x_n^{(g_k)}$ as \begin{equation}
 \mathbf{X}^{(g)} \triangleq \left[ \begin{matrix}
\mathbf{X}^{(g_1)} & \mathbf{X}^{(g_2)} &\hdots & \mathbf{X}^{(g_{K_g})}
 \end{matrix} \right]_{T \times K_g L } ,
\end{equation} where $\left( \mathbf{X}^{(g_k)} \right)_{ij} \triangleq x_{i-j}^{(g_k)} $ for $i=0,\dots,T-1$ and $j=0,\dots,L-1$.
% as
%\begin{equation}
%\mathbf{X}^{(g_k)} = \left[
%\begin{matrix}
%x_0^{(g_k)} & x_{-1}^{(g_k)} & \dots & x_{-(L-1)}^{(g_k)} \\
%x_1^{(g_k)} & x_0^{(g_k)} & \dots & x_{-(L-2)}^{(g_k)} \\
%\vdots & \vdots & \ddots & \vdots \\
%x_{T-1}^{(g_k)} & x_{T-2}^{(g_k)} & \dots & x_{-(T-L)}^{(g_k)} \\
%\end{matrix}
%\right]
%\end{equation} 
If there is a preamble cycle before training, symbols from the preamble; otherwise previously decoded data symbols are employed for positions with negative time indices. 

In the training period, only the intended group $\tilde{g}$ trains and other groups are in the data mode. Therefore, $\mathbf{X}^{(\tilde{g})}$ is a deterministic matrix while $\mathbf{X}^{(g)}$ for $g \neq \tilde{g}$ are random matrices obeying symbol properties described in Section (\ref{wsctr}). 

Estimated effective channel is found by \begin{equation} \label{ch_est_gen}
\hat{\bar{\mathbf{h}}}_{eff}^{(\tilde{g})} = \mathbf{Z}^{(\tilde{g})} \bar{\mathbf{s}}^{(\tilde{g})} ,
\end{equation} where $\mathbf{Z}^{(\tilde{g})}$ is the estimator matrix for the intended group $\tilde{g}$ and of size $K_{\tilde{g}} L D_{\tilde{g}} \times T D_{\tilde{g}}$. Then, the long vector $\hat{\bar{\mathbf{h}}}_{eff}^{(\tilde{g})}$ is partitioned to get needed effective channel estimates $\hat{\mathbf{h}}_{eff,l}^{(\tilde{g}_k)}$ according to the structure given in (\ref{h_eff_bar1}) and (\ref{h_eff_bar2}). 

\subsubsection{Reduced Rank Approximated Minimum Mean Square Error Estimator}
The following is a reduced-rank MMSE type estimator which is designed after analog beamforming \cite{StevenKay}. It is an approximated MMSE estimator since the CCMs in reduced-dimension are assumed to be perfectly acquired. It is given as
\begin{equation} \label{rr_mmse_est}
\mathbf{Z}^{(\tilde{g})} = \hat{\bar{\mathbf{C}}}^{(\tilde{g})} \left( \hat{\bar{\mathbf{R}}}_s^{(\tilde{g})} \right)^{-1} ,
\end{equation} where $\hat{\bar{\mathbf{R}}}_s^{(\tilde{g})}$ is the estimate for $\bar{\mathbf{R}}_s^{(\tilde{g})} \triangleq E\left\lbrace \bar{\mathbf{s}}^{(\tilde{g})} \left[ \bar{\mathbf{s}}^{(\tilde{g})} \right]^{H} \right\rbrace$ and $\hat{\bar{\mathbf{C}}}^{(\tilde{g})}$ is the estimate for $\bar{\mathbf{C}}^{(\tilde{g})} \triangleq  E\left\lbrace \bar{\mathbf{h}}_{eff}^{(\tilde{g})} \left[ \bar{\mathbf{s}}^{(\tilde{g})} \right]^{H} \right\rbrace$. $\bar{\mathbf{R}}_s^{(\tilde{g})}$ and $\bar{\mathbf{C}}^{(\tilde{g})}$ are given in (\ref{r_y_bar}) and (\ref{phi_bar}), respectively.
\begin{equation} \label{r_y_bar}
\bar{\mathbf{R}}_s^{(\tilde{g})} = \left( \mathbf{X}^{(\tilde{g})} \otimes \mathbf{I}_{D_{\tilde{g}}} \right) \bar{\mathbf{R}}_{eff}^{(\tilde{g})} \left( \mathbf{X}^{(\tilde{g})} \otimes \mathbf{I}_{D_{\tilde{g}}} \right)^{H} 
 + \mathbf{I}_T \otimes \mathbf{R}_{eff,\eta}^{(\tilde{g})}
\end{equation}
\begin{equation} \label{phi_bar}
\bar{\mathbf{C}}^{(\tilde{g})} = \bar{\mathbf{R}}_{eff}^{(\tilde{g})} \left( \mathbf{X}^{(\tilde{g})} \otimes \mathbf{I}_{D_{\tilde{g}}} \right)^{H}
\end{equation}

$\bar{\mathbf{R}}_{eff}^{(\tilde{g})} \triangleq  E\left\lbrace \bar{\mathbf{h}}_{eff}^{(\tilde{g})} \left[ \bar{\mathbf{h}}_{eff}^{(\tilde{g})} \right]^{H} \right\rbrace$ in (\ref{r_y_bar}) and (\ref{phi_bar}) can be found substituting $g=\tilde{g}$ in following expression.
\begin{equation} \label{r_eff_bar}
\bar{\mathbf{R}}_{eff}^{(g)} \triangleq \mathbf{I}_{K_{g}} \otimes \left( \sum_{l=0}^{L-1} \mathbf{E}_{L,l+1} \otimes \mathbf{R}_{eff,l}^{(g)} \right)
\end{equation}

In (\ref{r_eff_bar}), 
\begin{equation}
\mathbf{E}_{L,l+1} \triangleq \mathbf{e}_{L,l+1} \left(\mathbf{e}_{L,l+1} \right)^{H},
\end{equation} where $\mathbf{e}_{L,l+1}$ is the $(l+1)$\textsuperscript{st} column of $\mathbf{I}_L$. $\hat{\bar{\mathbf{R}}}_s^{(\tilde{g})}$ and $\hat{\bar{\mathbf{C}}}^{(\tilde{g})}$ in \eqref{rr_mmse_est} are obtained using $\hat{\mathbf{R}}_{eff,l}^{(\tilde{g})}$ and $\hat{\mathbf{R}}_{eff,\eta}^{(\tilde{g})}$ (which are described in Section \ref{recproc}) instead of $\mathbf{R}_{eff,l}^{(\tilde{g})}$ and $\mathbf{R}_{eff,\eta}^{(\tilde{g})}$ in (\ref{r_y_bar}) and (\ref{r_eff_bar}).

\subsubsection{Analytical Expression for Normalized Mean Squared Error}
Normalized MSE will be used as a measure of instantaneous channel acquisition performance. For a slow-time instant, normalized MSE is defined as
\begin{equation}
{nMSE}^{(\tilde{g})} = \frac{E_{\mathbf{Z}^{(\tilde{g})},\mathbf{S}^{(\tilde{g})}} \left[ E\left[ \left|| \bar{\mathbf{h}}_{eff}^{(\tilde{g})} - \hat{\bar{\mathbf{h}}}_{eff}^{(\tilde{g})} |\right|^2 | \, \mathbf{Z}^{(\tilde{g})},\mathbf{S}^{(\tilde{g})} \right] \right]}{E_{\mathbf{S}^{(\tilde{g})}} \left[ E\left[ \left|| \bar{\mathbf{h}}_{eff}^{(\tilde{g})} |\right|^2 | \, \mathbf{S}^{(\tilde{g})} \right] \right]}
\end{equation}

%\begin{equation}
%E\left[ \left|| \bar{\mathbf{h}}_{eff}^{(g)} - \hat{\bar{\mathbf{h}}}_{eff}^{(g)} |\right|^2 | \, \mathbf{Z}^{(g)},\mathbf{S}^{(g)} \right] = \text{tr} \left\lbrace \mathbf{R}_{mmse}^{(g)} \right\rbrace + \text{tr} \left\lbrace \left( \mathbf{W}^{(g)} - \mathbf{W}_{mmse}^{(g)} \right)^{H} \mathbf{R}_{y}^{(g)} \left( \mathbf{W}^{(g)} - \mathbf{W}_{mmse}^{(g)} \right) \right\rbrace
%\end{equation}
 Numerator and denominator can be expressed analytically as
\begin{align}
E\left[ \left|| \bar{\mathbf{h}}_{eff}^{(\tilde{g})} - \hat{\bar{\mathbf{h}}}_{eff}^{(\tilde{g})} |\right|^2 | \, \mathbf{Z}^{(\tilde{g})},\mathbf{S}^{(\tilde{g})} \right] & = \text{tr} \left\lbrace \bar{\mathbf{R}}_{eff}^{(\tilde{g})} \right\rbrace +
\text{tr} \left\lbrace \mathbf{\Phi} \right\rbrace -
2Re\left\lbrace \text{tr} \left\lbrace \mathbf{\Psi}^{(\tilde{g})} \right\rbrace \right\rbrace
\\
E\left[ \left|| \bar{\mathbf{h}}_{eff}^{(\tilde{g})} |\right|^2 | \, \mathbf{S}^{(\tilde{g})} \right] & = \text{tr} \left\lbrace \bar{\mathbf{R}}_{eff}^{(\tilde{g})} \right\rbrace,
\end{align}
where $\mathbf{\Phi} \triangleq E \left\lbrace \hat{\bar{\mathbf{h}}}_{eff}^{(\tilde{g})} \left[ \hat{\bar{\mathbf{h}}}_{eff}^{(\tilde{g})} \right]^{H} \right\rbrace$ and $\mathbf{\Psi}^{(g)} \triangleq E \left\lbrace \hat{\bar{\mathbf{h}}}_{eff}^{(\tilde{g})} \left[  \bar{\mathbf{h}}_{eff}^{(g)} \right]^{H} \right\rbrace$ are found to be as in (\ref{phi}) and (\ref{psi}), respectively; using (\ref{ch_est1}) , (\ref{ch_est_gen}) and the fact that $E\left\lbrace \bar{\mathbf{h}}_{eff}^{(g)} \left[ \bar{\mathbf{h}}_{eff}^{(g')} \right]^{H} \right\rbrace = \delta_{gg'} \bar{\mathbf{R}}_{eff}^{(g)}$.
\begin{align} \label{phi}
\mathbf{\Phi} & = \mathbf{Z}^{(\tilde{g})} \bar{\mathbf{R}}_{s}^{(\tilde{g})} \left[ \mathbf{Z}^{(\tilde{g})} \right]^{H}
\\
 \label{psi}
\mathbf{\Psi}^{(g)} & = \delta_{g \tilde{g}} \mathbf{Z}^{(\tilde{g})} \left( \mathbf{X}^{(g)} \otimes \mathbf{I}_{D_{\tilde{g}}} \right) \bar{\mathbf{R}}_{eff}^{(g)}
\end{align}

\subsection{Asymptotic Analysis for the Recursive Filtering Output}\label{asymptotic}
Consider that there is only one user in the environment and it has only one MPC. Let its channel be static such that the center angle $\mu_{\theta}$ and the angular spread $\sigma_{\theta}$ does not change with time. However, the estimate $\hat{\mu}_{\theta}[n]$ for the $n$\textsuperscript{th} time instance is erroneous such that
\begin{equation}
\hat{\mu}_{\theta}[n]=\mu_{\theta} + e_{\theta}[n],
\end{equation}
where $e_{\theta}[n] \sim \mathcal{N}\left( 0 , \sigma_e^2 \right)$ and i.i.d. for different $n$ (Although $e_{\phi}[n] \sim \mathcal{N}\left( 0 , \sigma_{est}^2 \right)$, $e_{\theta}[n]$ and $e_{\phi}[n]$ can be assumed to be linearly related for small $\sigma_{est}$ values considered in this work; see Appendix for transformation in \eqref{transformation}). In addition, the angular spread estimate is perfect such that $\hat{\sigma}_{\theta}[n]=\sigma_{\theta}$. The estimate CCM $\hat{\mathbf{R}}[n]$ is constructed according to values $\hat{\mu}_{\theta}[n]$ and $\sigma_{\theta}$. The estimate CCM $\hat{\mathbf{R}}[n]$ can be expressed as
\begin{equation}
\hat{\mathbf{R}}[n] = \mathbf{R} \odot \left( \mathbf{q}(e_{\theta}[n]) \mathbf{q}^{H}(e_{\theta}[n]) \right),
\end{equation}
using a similar approach to that applied for \eqref{eqhadamardfac}, meaning its entries are
\begin{equation}
\left( \hat{\mathbf{R}} [n] \right)_{ab} = \left( \mathbf{R} \right)_{ab} e^{j(a-b) e_{\theta}[n]}.
\end{equation}

The mean, variance and cross correlation of the exponential term, with a normally distributed random variable in its exponent, can be expressed with the help of characteristic function for the normal random variables as shown below.
\begin{equation}
E\left[ e^{j(a-b) e_{\theta}[n]} \right] = e^{-\frac{1}{2}(a-b)^2 \sigma_e^2}
\end{equation}
\begin{equation}
\text{var}\left( e^{j(a-b) e_{\theta}[n]} \right) = 1 - e^{-(a-b)^2 \sigma_e^2}
\end{equation}
\begin{align}
E & \left[ \left( e^{j(a-b) e_{\theta}[n_1]} -E\left[ e^{j(a-b) e_{\theta}[n_1]} \right] \right) \left( e^{j(a-b) e_{\theta}[n_2]} -E\left[ e^{j(a-b) e_{\theta}[n_2]} \right] \right)^* \right] \nonumber \\
& = E \left[ e^{j(a-b) ( e_{\theta}[n_1] - e_{\theta}[n_2])} \right] - \left( e^{-\frac{1}{2}(a-b)^2 \sigma_e^2} \right)^2 \\
&= e^{-\frac{1}{2}(a-b)^2 2\sigma_e^2} - e^{-(a-b)^2 \sigma_e^2} \\
&= 0 \, \, \text{for} \, \, n_1 \neq n_2
\end{align}

Then, the mean and variance of the entries of estimated CCM $\hat{\mathbf{R}} [n]$ are as below.
\begin{equation}
E\left[ \left( \hat{\mathbf{R}} [n] \right)_{ab} \right] = \left( \mathbf{R} \right)_{ab} e^{-\frac{1}{2}(a-b)^2 \sigma_e^2}
\end{equation}
\begin{equation}
\text{var} \left( \left( \hat{\mathbf{R}} [n] \right)_{ab} \right) = \left( \left( \mathbf{R} \right)_{ab} \right)^2 \left( 1 - e^{-(a-b)^2 \sigma_e^2} \right)
\end{equation}

The output of the recursive filtering operation can be expressed as
\begin{align}
\hat{\mathbf{R}}^{f}[n] &= \beta \, \hat{\mathbf{R}}^{f}[n-1] + \left(1-\beta\right)\hat{\mathbf{R}}[n] \\
&= \beta^n \hat{\mathbf{R}}[0] + (1-\beta) \beta^n \sum_{k=1}^n \beta^{-k} \hat{\mathbf{R}}[k]
\end{align}

The entries of the output are
\begin{equation}
\left( \hat{\mathbf{R}}^{f}[n] \right)_{ab} = \beta^n \left( \mathbf{R} \right)_{ab} e^{j(a-b) e_{\theta}[0]} + (1-\beta) \beta^n \sum_{k=1}^n \beta^{-k} \left( \mathbf{R} \right)_{ab} e^{j(a-b) e_{\theta}[k]} 
\end{equation}

Using the mean, variance and uncorrelatedness of the exponential random term which are found above, the mean and the variance of the entries of the filtered output CCM are calculated as below.
\begin{align}
E \left[ \left( \hat{\mathbf{R}}^{f}[n] \right)_{ab} \right] &= \beta^n \left( \mathbf{R} \right)_{ab} e^{-\frac{1}{2}(a-b)^2 \sigma_e^2}  + (1-\beta) \beta^n \sum_{k=1}^n \beta^{-k} \left( \mathbf{R} \right)_{ab} e^{-\frac{1}{2}(a-b)^2 \sigma_e^2} \\
&=  \left( \mathbf{R} \right)_{ab} e^{-\frac{1}{2}(a-b)^2 \sigma_e^2} \label{meanccmrecfilt}
\end{align}

\begin{align}
\text{var} \left( \left( \hat{\mathbf{R}}^{f}[n] \right)_{ab} \right) &= \beta^{2n} \left( \mathbf{R} \right)_{ab}^2 \left(1 - e^{-(a-b)^2 \sigma_e^2} \right) \nonumber \\
& \quad + (1-\beta)^2 \beta^{2n} \sum_{k=1}^n \beta^{-2k} \left(\left( \mathbf{R} \right)_{ab}\right)^2 \left(1 - e^{-(a-b)^2 \sigma_e^2} \right) \\
&= \left( \mathbf{R} \right)_{ab}^2 \left(1 - e^{-(a-b)^2 \sigma_e^2} \right) \left( \beta^{2n} + \frac{1-\beta}{1+\beta} \left(1-\beta^{2n}\right)\right)
\end{align}

\begin{equation}
\lim_{n \rightarrow \infty} \text{var} \left( \left( \hat{\mathbf{R}}^{f}[n] \right)_{ab} \right) = \left( \mathbf{R} \right)_{ab}^2 \left(1 - e^{-(a-b)^2 \sigma_e^2} \right)\frac{1-\beta}{1+\beta}
\end{equation}

\begin{equation}
\lim_{n \rightarrow \infty} \text{var} \left( \left( \hat{\mathbf{R}}^{f}[n] \right)_{ab} \right) = \frac{1-\beta}{1+\beta} \text{var} \left( \left( \hat{\mathbf{R}}[n] \right)_{ab} \right)
\end{equation}

\begin{figure}[h]
\centerline{\includegraphics[width=0.5\textwidth]{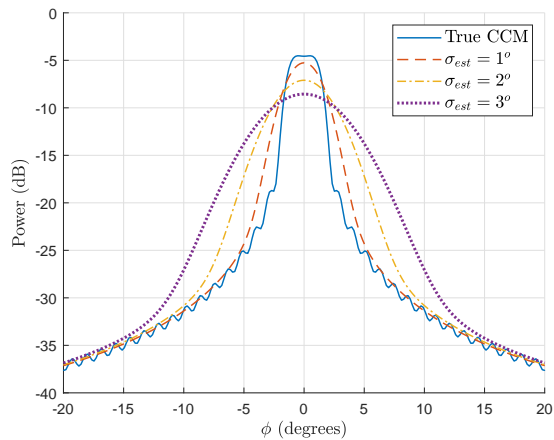}}
\caption{The spreading effect of recursive filtering with erroneous estimates}
\label{fig_ryf_spread}
\end{figure}

As it is seen, the variance decreases with filtering with a mean that is not the same as the actual CCM. The new mean of the filtered output CCM has wider spread in angular domain than the actual CCM. The spreading effect of recursive filtering with erroneous estimates is shown in Figure \ref{fig_ryf_spread} where 
\begin{equation}
\mathbf{u}^{H}(\phi) E \left[ \hat{\mathbf{R}}^{f}[n] \right] \mathbf{u}(\phi)\end{equation} 
is plotted using (\ref{meanccmrecfilt}) for various error levels.

Although the mean calculated for the lastly estimated CCM is equal to the that calculated for the filtered one, it is the mean of the process. A single realization of this process is used every time which carries a biased mean AoA estimation. Therefore, while the usage of the last estimate CCM is related with biased estimation, the usage of recursively filtered estimate CCM is related with a spread error. In the previous work \cite{kurt19}, it was shown that the system is vulnerable against the biased angular estimations but an error in the estimation of angular spread can be tolerated. Therefore, constructing GEB with recursively filtered CCMs promises a performance increase when the angular estimates are biased. However, the recursive filtering operation does not result in a convergence to the actual matrix. Therefore, there might be cases in which its usage is disadvantageous.

\section{Numerical Results}\label{secresults}
In this section, some numerical results will be given using the scenario in Table \ref{tb_angular_sec2}, for intended group $\tilde{g}=1$, one RF chain for each MPC ($D_1=3 \times 1 =3$), perfect instantaneous channel estimation ($\hat{\mathbf{H}}_{eff,l}^{(g)} = \mathbf{H}_{eff,l}^{(g)}$ but $\hat{\mathbf{R}}_{l}^{(g)} \neq \mathbf{R}_{l}^{(g)}$), 30 dB input SNR, $\sigma_{v}=3^o$ asymptotic AoA standard deviation, and $N=100$, unless otherwise is stated.
\begin{table}[b]
\caption{Angular sectors of multipath components}
\begin{center}
\resizebox{0.6\textwidth}{!}{%
\begin{tabular}{|c|c|c|c|c|c|c|c|}
\hline
%\textbf{Group Index ($g$)} & \textbf{Center Angle ($\mu_{\phi}$)} & \textbf{Angular Spread ($\sigma_{\phi}$)} & \textbf{Delay ($\mathcal{L}^{(g)}[m]$)} & \textbf{MPC Power Distribution} & \textbf{RF Chain ($d_m^{(g)}$)} & \textbf{Number of Users ($K_g$)} & \textbf{Signal Power ($E_s^{(g)}$)}\\
\textbf{Group} & \textbf{N. of} & \textbf{Center} & \textbf{Angular} & \textbf{RF} & \textbf{N. of} & \textbf{Signal}\\
\textbf{Index} & \textbf{MPCs} & \textbf{Angle} & \textbf{Spread} & \textbf{Delay} & \textbf{Users} & \textbf{Power}\\
\textbf{($g$)} & \textbf{($M^{(g)}$)} & \textbf{($\mu_{\phi}$)} & \textbf{($\sigma_{\phi}$)} & \textbf{($\mathcal{L}^{(g)}(m)$)} & \textbf{($K_g$)} & \textbf{($E_s^{(g)}$)}\\
\hline \hline
1 & 3 & (0\textdegree, 9.75\textdegree, 22\textdegree) & (3\textdegree,   2.5\textdegree, 3.5\textdegree) & (0, 5, 11) & 1 & 1  \\
\hline
2 & 2 & (27.5\textdegree, 15.25\textdegree) & (3\textdegree, 2\textdegree) & (3, 9) & 2 & 10 \\
\hline
3 & 2 & (-6.25\textdegree, -13.5\textdegree) & (3.5\textdegree, 3\textdegree) & (8, 17) & 3 & 100 \\
\hline
4 & 2 & (-20.25\textdegree, -27\textdegree) & (3\textdegree, 3.5\textdegree) & (20, 29) & 4 & 1000 \\
\hline
\end{tabular}
\label{tb_angular_sec2}
}
\end{center}
\end{table}  

\begin{figure}[h]
\centerline{\includegraphics[width=0.60\textwidth]{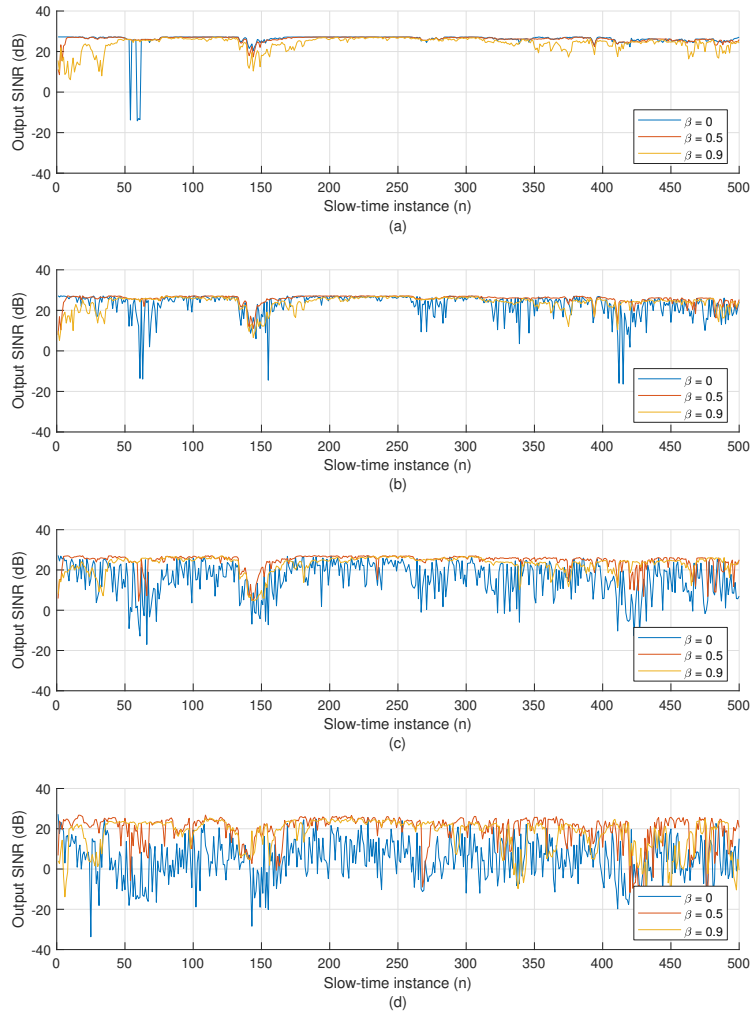}}
\caption{The performance of the recursive filtering combined with GEB versus slow-time for the mobile channel with $\alpha=0.999$. (a) $\sigma_{est}=0.1$. (b) $\sigma_{est}=0.5$. (c) $\sigma_{est}=1$. (d) $\sigma_{est}=2$.}
\label{fig_sinrvstime}
\end{figure}

 In Figure \ref{fig_sinrvstime}, the performance of the GEB method combined with the recursive filtering is given. The recursive filtering approach is actively used in the cases of $\beta=0.5$ and $\beta=0.9$, but not in the case of $\beta=0$. It is seen that the bias in the angular estimation seriously affects the performance for $\beta=0$, while recursive filtering brings a remarkable robustness. 

%\begin{figure}[h]
%\centerline{\includegraphics[width=0.45\textwidth]{alpha4.eps}}
%\caption{Performance against channel mobility index $\alpha$. (a) $\sigma_{est}=0.1$. (b) $\sigma_{est}=0.5$. (c) $\sigma_{est}=1$. (d) $\sigma_{est}=2$.}
%\label{fig_alpha}
%\end{figure}
 
\begin{figure}[h]
\centerline{\includegraphics[width=0.70\textwidth]{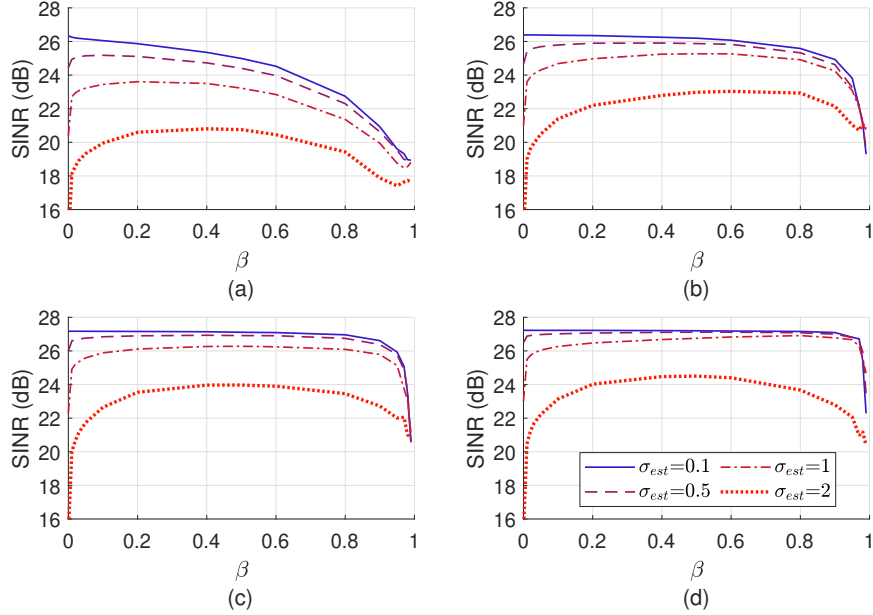}}
\caption{Performance against recursive filtering parameter $\beta$. (a) $\alpha=0.9$. (b) $\alpha=0.99$. (c) $\alpha=0.999$. (d) $\alpha=0.9999$.}
\label{fig_beta}
\end{figure}

 The next figures will show the slow-time-averaged performances. The change of performance against the recursive filtering parameter $\beta$ is given in Figure \ref{fig_beta}. When subplots are compared focusing on the more accurate angular estimate curves with $\sigma_{est}=0.1$, it is seen that the usage of recursive filtering in fast channels (with small $\alpha$) decreases the performance, while the parameter $\beta$ does not differ the performance in a wide range for slow channels. When the angular estimation accuracy decreases (as $\sigma_{est}$ increases), an optimum value for the $\beta$ parameter starts to be apparent in related curves. For $\sigma_{est}=2$, the optimum value is the most apparent and it is about $\beta=0.5$. 

%\begin{figure}[h]
%\centerline{\includegraphics[width=0.45\textwidth]{beta4.jpg}}
%\caption{Performance against filtering parameter $\beta$ ((a) : $\sigma_{est}=0.1$, (b) : $\sigma_{est}=0.5$, (c) : $\sigma_{est}=1$, (d) : $\sigma_{est}=2$)}
%\label{fig_beta2}
%\end{figure}

%The change of performance against the standard deviation of the angular estimation error $\sigma_{est}$ is given in Figure \ref{fig_sigma_est}. Recursive filtering is seen to bring a robustness against angular estimation errors. It should be avoided only when channel is fast and the angular estimation is successful. Comparing the subplots, it is seen that the performance generally drops as the mobility gets faster. 
%
%\begin{figure}[h]
%\centerline{\includegraphics[width=0.70\textwidth]{sigma_est8_exp.eps}}
%\caption{Performance against standard deviation of the channel estimation error $\sigma_{est}$. (a) $\alpha=0.9$. (b) $\alpha=0.99$. (c) $\alpha=0.999$. (d) $\alpha=0.9999$.}
%\label{fig_sigma_est}
%\end{figure}
% 
\begin{figure}[h]
\centerline{\includegraphics[width=0.70\textwidth]{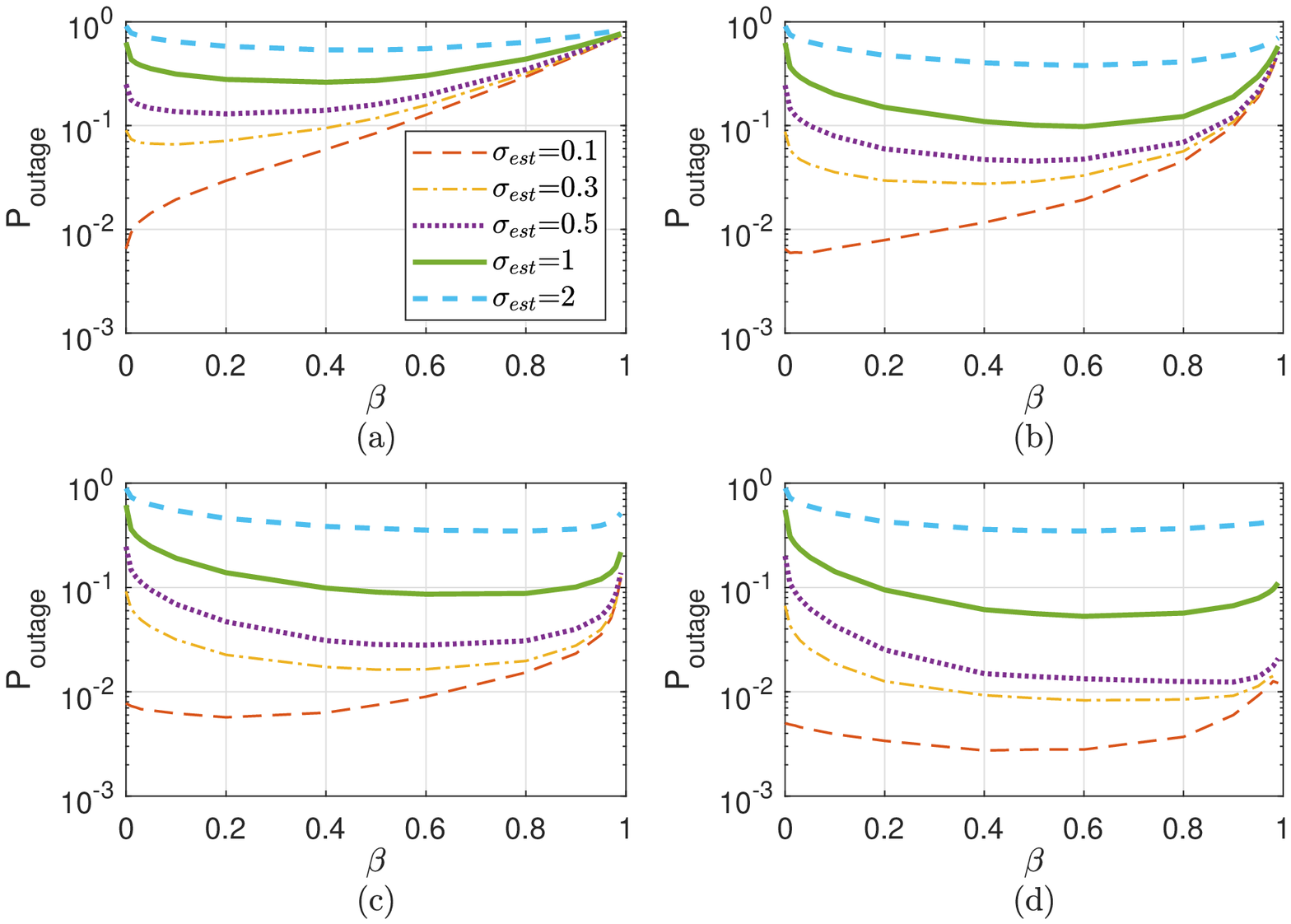}}
\caption{Change of probability of outage with recursive filtering parameter $\beta$. (a) $\alpha=0.9$. (b) $\alpha=0.99$. (c) $\alpha=0.999$. (d) $\alpha=0.9999$.}
\label{fig_outage_alpha}
\end{figure}
%\begin{figure}[h]
%\centerline{\includegraphics[width=0.45\textwidth]{outage_sigma_est3.eps}}
%\caption{Change of probability of outage with recursive filtering parameter $\beta$. (a) $\sigma_{est}=0.1$. (b) $\sigma_{est}=0.5$. (c) $\sigma_{est}=1$. (d) $\sigma_{est}=2$.}
%\label{fig_outage_sigma_est}
%\end{figure}

Experiments for probability of outage were conducted for Figure \ref{fig_outage_alpha}, in which input SNR is held at 30 dB and slow-time-expected output SINR below 20 dB is assumed as outage. It is seen that outage due to high channel estimation error can be decreased via proposed recursive filtering scheme. However, the presence of an optimum value should be noted. In addition, in very small channel estimation error cases, it is seen that the proposed method should be avoided.
\begin{figure}[h]
\centerline{\includegraphics[width=0.60\textwidth]{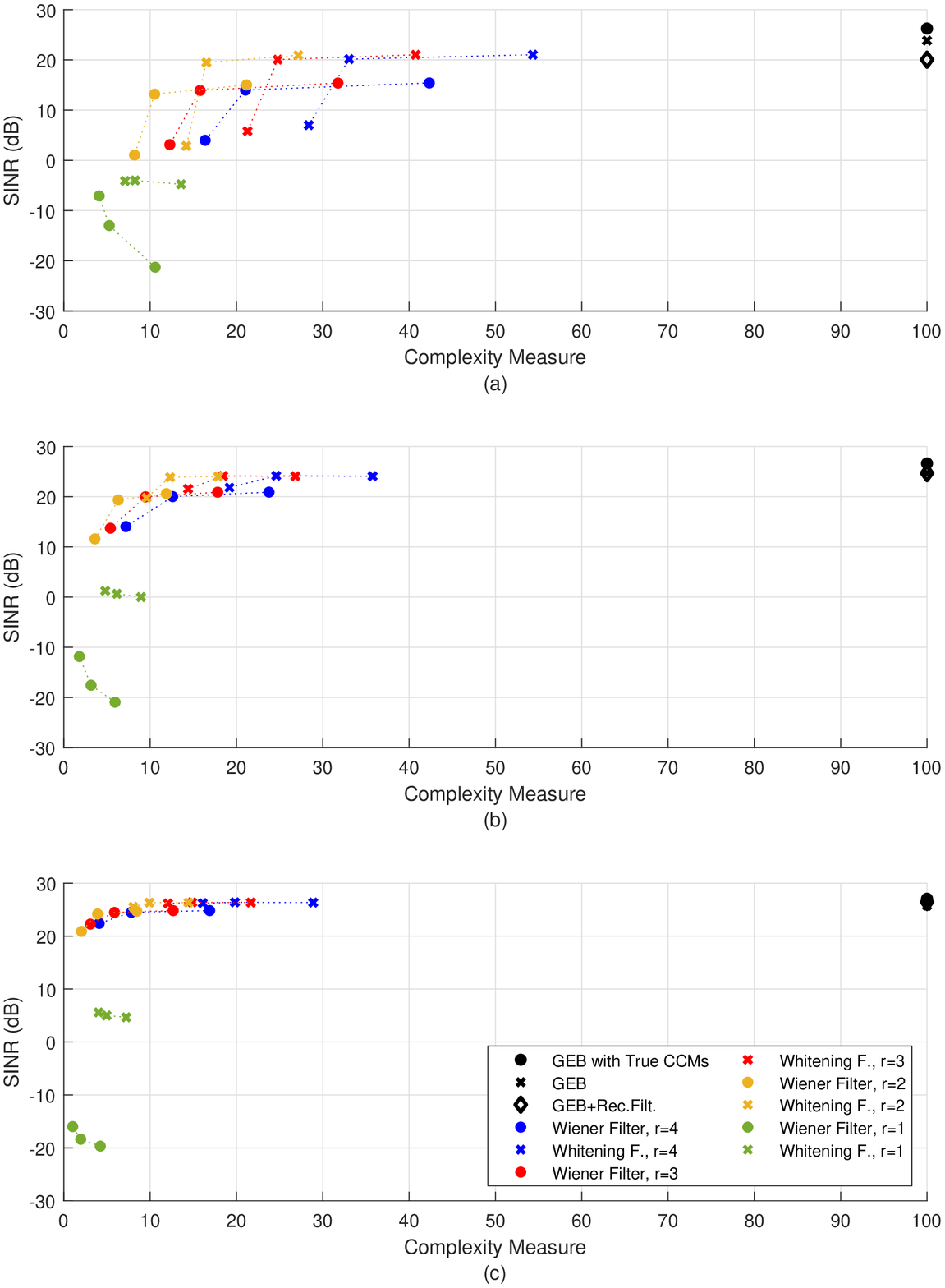}}
\caption{Comparison of complexity and performance of different methods for $\beta=0.9$ and $\sigma_{est}=0.5$. (a) $\alpha=0.9$. (b) $\alpha=0.99$. (c) $\alpha=0.999$.}
\label{fig_complexitya}
\end{figure}

%\begin{figure}[h]
%\centerline{\includegraphics[width=0.45\textwidth]{comp_b.eps}}
%\caption{Comparison of complexity and performance of different methods for $\alpha=0.99$ and $\sigma_{est}=0.5$. (a) $\beta=0$. (b) $\beta=0.5$. (c) $\beta=0.9$.}
%\label{fig_complexityb}
%\end{figure}
%
%\begin{figure}[h]
%\centerline{\includegraphics[width=0.45\textwidth]{comp_s.eps}}
%\caption{Comparison of complexity and performance of different methods for $\alpha=0.99$ and $\beta=0.9$. (a) $\sigma_{est}=0.1$. (b) $\sigma_{est}=0.5$. (c) $\sigma_{est}=2$.}
%\label{fig_complexitys}
%\end{figure}

For the complexity analysis, complexity measure is calculated as $N_{\Delta p} r$ and $(\, N_{\Delta p} + N_{p,l}^{(g)} \,)r$ for the Wiener filter type and whitening filter type methods, respectively. $r$ is a design variable. A Monte Carlo analysis performed for $N_{\Delta p}$, which is dependent on the mobility of the channel. The results are given in Table \ref{tablechanges2q}. It is seen that the change in the quantized patch power levels are greater with faster channels, larger estimation errors and more detailed quantization, as expected. On the other hand, the usage of the recursive filtering with large $\beta$ parameter is seen to decrease the change remarkably. However, this decrease is beneficial only if it does not bring a significant decrease in the performance.
%
%\begin{table}[b]
%\caption{Average number of changes in the quantized patch power levels for quantization parameter k=1}
%\begin{center}
%\resizebox{0.8\textwidth}{!}{%
%\begin{tabular}{|c|c|c|c|c|c|c|c|c|c|c|c|c|}
%\hline
% {$\alpha$} & \multicolumn{3}{|c|}{ {0.9}} & \multicolumn{3}{|c|}{ {0.99}} & \multicolumn{3}{|c|}{ {0.999}} & \multicolumn{3}{|c|}{ {0.9999}} \\
% \hline
% $\beta$ & 0 & 0.5 & 0.9 & 0 & 0.5 & 0.9 & 0 & 0.5 & 0.9 & 0 & 0.5 & 0.9 \\
% \hline \hline
%$\sigma_{est}=0$ & 14.84 & 12.46 & 3.93 & 4.98 & 3.96 & 1.5 & 1.59 & 1.21 & 0.48 & 0.52 & 0.4 & 0.16 \\
%\hline
%$\sigma_{est}=0.1$ & 14.78 & 12.57 & 3.97 & 5.19 & 4.15 & 1.55 & 2.29 & 1.63 & 0.5 & 1.77 & 1.15 & 0.26 \\
%\hline
%$\sigma_{est}=0.3$ & 15.36 & 12.95 & 4.03 & 6.9 & 5.3 & 1.63 & 5.16 & 3.63 & 0.71 & 4.89 & 3.43 & 0.56\\
%\hline
%$\sigma_{est}=0.5$ & 16.51 & 13.86 & 4.09 & 9.32 & 7.21 & 1.8 & 8.17 & 5.94 & 1.02 & 8.22 & 5.88 & 0.9\\
%\hline
%$\sigma_{est}=1$ & 20.35 & 16.89 & 4.44 & 16.16 & 12.77 & 2.58 & 15.67 & 12 & 1.93 & 15.68 & 12.35 & 1.88\\
%\hline
%$\sigma_{est}=2$ & 27.62 & 22.17 & 4.51 & 26.44 & 21.03 & 4.88 & 26.2 & 21.09 & 4.62 & 26.73 & 21.31 & 4.52\\
%\hline
%\end{tabular}
%}
%\label{tablechanges1q}
%\end{center}
%\end{table}
\begin{table}[b]
\caption{Average number of changes in the quantized patch power levels for quantization parameter $N_q$=2}
\begin{center}
\resizebox{0.5\textwidth}{!}{%
\begin{tabular}{|c|c|c|c|c|c|c|c|c|c|}
\hline
 {$\alpha$} & \multicolumn{3}{|c|}{ {0.9}} & \multicolumn{3}{|c|}{ {0.99}} & \multicolumn{3}{|c|}{ {0.999}} \\
 \hline
 $\beta$ & 0 & 0.5 & 0.9 & 0 & 0.5 & 0.9 & 0 & 0.5 & 0.9 \\
 \hline \hline
%$\sigma_{est}=0$ & 15.42 & 17.14 & 5.18 & 5.77 & 6.35 & 2.64 & 1.95 & 1.86 & 0.86 \\
%\hline
$\sigma_{est}=0.1$ & 15.36 & 17.18 & 5.13 & 5.96 & 6.58 & 2.64 & 2.8 & 2.43 & 0.97 \\
\hline
%$\sigma_{est}=0.3$ & 15.89 & 17.6 & 5.17 & 7.71 & 7.78 & 2.87 & 5.93 & 5 & 1.4 \\
%\hline
$\sigma_{est}=0.5$ & 17.02 & 18.28 & 5.26 & 10.06 & 9.6 & 3.15 & 8.95 & 7.62 & 1.96 \\
\hline
$\sigma_{est}=1$ & 20.78 & 20.62 & 5.39 & 16.69 & 14.98 & 4.05 & 16.22 & 13.72 & 3.32 \\
\hline
$\sigma_{est}=2$ & 27.87 & 25.74 & 6.75 & 26.73 & 23.71 & 4.87 & 26.48 & 23.63 & 4.72 \\
\hline
\end{tabular}
}
\label{tablechanges2q}
\end{center}
\end{table}

%\begin{table}[b]
%\caption{Average number of changes in the quantized patch power levels for quantization parameter k=4}
%\begin{center}
%\resizebox{0.8\textwidth}{!}{%
%\begin{tabular}{|c|c|c|c|c|c|c|c|c|c|c|c|c|}
%\hline
% {$\alpha$} & \multicolumn{3}{|c|}{ {0.9}} & \multicolumn{3}{|c|}{ {0.99}} & \multicolumn{3}{|c|}{ {0.999}} & \multicolumn{3}{|c|}{ {0.9999}} \\
% \hline
% $\beta$ & 0 & 0.5 & 0.9 & 0 & 0.5 & 0.9 & 0 & 0.5 & 0.9 & 0 & 0.5 & 0.9 \\
% \hline \hline
%$\sigma_{est}=0$  & 21.79 & 26.96 & 10.35 & 13.42 & 15.07 & 5.21 & 5.62 & 6.29 & 2.46 & 2.06 & 2.22 & 0.86 \\
%\hline
%$\sigma_{est}=0.1$ & 21.75 & 27.07 & 10.29 & 13.58 & 15.45 & 5.29 & 7.68 & 8.44 & 2.59 & 6.25 & 6.42 & 1.49  \\
%\hline
%$\sigma_{est}=0.3$ & 22.13 & 27.43 & 10.37 & 15.46 & 17.15 & 5.53 & 13.83 & 14.45 & 3.54 & 13.75 & 14.12 & 3.02 \\
%\hline
%$\sigma_{est}=0.5$ & 22.92 & 28.34 & 10.58 & 17.35 & 19.29 & 5.94 & 16.32 & 17.87 & 4.22 & 17.65 & 17.79 & 3.86 \\
%\hline
%$\sigma_{est}=1$ & 25.91 & 31.38 & 11.19 & 22.91 & 25.49 & 7.41 & 22.52 & 23.6 & 6.19 & 22.51 & 25.19 & 6.15 \\
%\hline
%$\sigma_{est}=2$  & 31.4 & 37.69 & 12.65 & 30.49 & 35.3 & 10.87 & 30.21 & 35.6 & 10.56 & 31.22 & 36.04 & 10.55 \\
%\hline
%\end{tabular}
%}
%\label{tablechanges4q}
%\end{center}
%\end{table}

For the whitening filter type method, $N_{p,l}^{(g)}$ is taken as 3, and $\hat{\sigma}_{\theta,l}^{(g)}=3^{o}$ is used for $\hat{\mathbf{w}}_{1,l}^{(g)}[n]$ in \eqref{gevecRlg}, considering the angular spreads in the scenario. Scatter plots in Figure \ref{fig_complexitya} are prepared for the comparison of complexity and performance of the proposed methods. Each marker indicates a beamformer type designed with a selected set of parameters. Marker triplets of same shape and color have the same effective rank $r$ for the matrix $\mathbf{D}$, for Wiener and whitening filter types. The three markers in those triplets differ in the quantization depth $N_q$ given in (\ref{quantization2}) and (\ref{quantization3}). From leftmost to rightmost, $N_q=1,2$ and $4$, respectively. For GEB methods, the complexity is taken as $N=100$, as explained before. 

In Figure \ref{fig_complexitya}, there are three plots for 0.9, 0.99 and 0.999 values of channel mobility rate parameter $\alpha$. The left and upper part of the plots is the effective region. It is seen that $r=1$ has a poor performance. When $r$ change is followed for 2,3 and 4 for the same filter type; in other words, when yellow, red and blue markers are followed for the same marker type, it is seen that red and blue markers are increasing in terms of complexity but stay almost the same in terms of performance. Therefore, $r=2$ seems as the optimum selection in terms of efficiency. When Wiener and whitening filters are observed for the same $r$ and quantization depth $N_q$; in other words, dots and crosses are compared for the same color and same position in the triplets, it is seen that the whitening Filter increases performance by increasing the complexity. Positions of the GEB methods and the cluster of proposed adaptive methods show that the proposed methods promises remarkable decrease in the complexity while not losing much from the performance.

Looking to graphs (a), (b) and (c) in the same figure, it is seen that when the channel mobility increases, adaptive methods experience a decrease in the performance and an increase in the complexity, and vice versa. There are some markers whose vertical decrease is limited, for example the whitening filter with $r=2$ and $N_q=2$. It is 4 dB below the best GEB solution that works with estimated CCMs even in the fastest case, which is GEB without recursive filtering indicated with black cross, while still providing a significant complexity decrease. In other cases, its output SINR loss is around 1 dB. It is seen from Figure \ref{fig_complexitya} that the GEB method for the analog beamformer design, whose advantages are mentioned in the previous work \cite{kurt19}, can be replaced by a modified, lower complexity adaptive method without losing from its performance level. 

%In Figure \ref{fig_complexityb}, the effect of the recursive filtering parameter $\beta$ is shown. It is seen that small $\beta$ values significantly increases the complexity measure, which is related with the change rate of the patch power levels. It was shown in previous figures that $\beta$ parameter has an apparent optimum value only for some cases, which is about 0.5. For most of the cases, there is a wide range of $\beta$ values that can be selected without experiencing a significant difference in the performance. Then, for the complexity of the adaptive methods, it is concluded that the $\beta$ value should be chosen around 0.9 if there is no reason to do otherwise.
%
%The effect of erroneous AoA estimations is seen in Figure \ref{fig_complexitys}. It is seen that it results in both a decrease in the performance and increase in the complexity. But the whitening filter with $r=2$ and $k=2$ seems to preserve the performance given by the best GEB method that works with the estimated CCMs, while providing a low complexity measure around 16, for the case in which $\sigma_{est}=2$.

\begin{figure}[h]
\centerline{\includegraphics[width=0.70\textwidth]{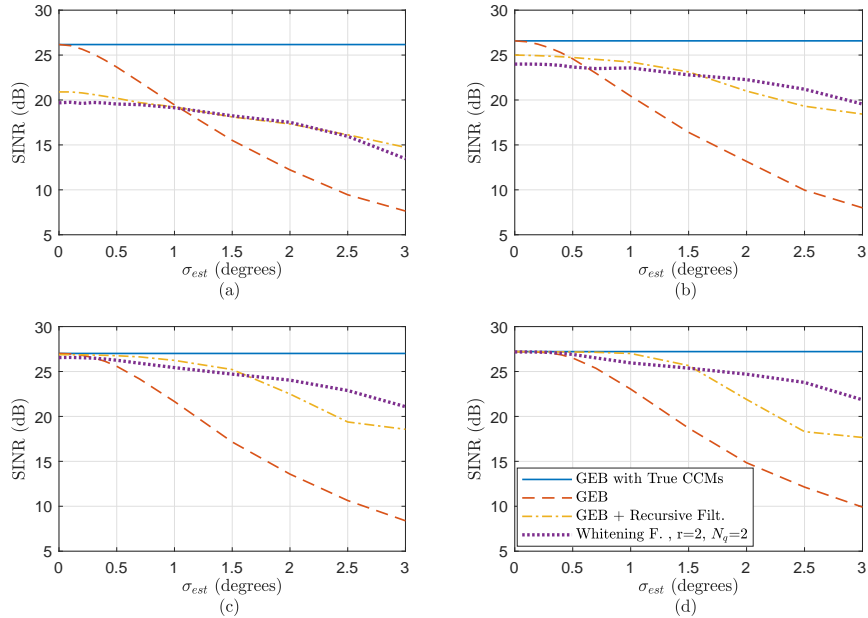}}
\caption[caption]{Output SINR performances of different methods versus $\sigma_{est}$; for group 1, $\beta=0.9$. (a) $\alpha=0.9$. (b) $\alpha=0.99$. (c) $\alpha=0.999$. (d) $\alpha=0.9999$.}
\label{fig_SINR_sigmaest_1}
\end{figure}

The change of performance against the standard deviation of the angular estimation error $\sigma_{est}$ is given in Figure \ref{fig_SINR_sigmaest_1}. There are 4 subplots with different channel mobilities, each of which includes 4 curves. In the first three, GEB is applied with true CCMs, estimated CCMs ($\beta=0$ case), and recursively filtered estimated CCMs with $\beta=0.9$. The fourth one belongs to the whitening filter with $r=2$, $N_q=2$ and $\beta=0.9$. Recursive filtering is seen to bring a robustness against angular estimation errors. It should be avoided only when channel is fast and the angular estimation is successful. It is seen that whitening filter can outperform the GEB with recursive filtering, although former is derived as an approximation to latter. It is due to the fact that the whitening filter implements a wait-and-see process against unreliable estimations through patch power quantization after recursive filtering, hence does not update in each slow-time instant, and is affected less by the effect mentioned in Section \ref{asymptotic}. Finally, comparing the subplots, it is seen that the performance generally drops as the mobility gets faster.

\begin{figure}[h]
\centerline{\includegraphics[width=0.70\textwidth]{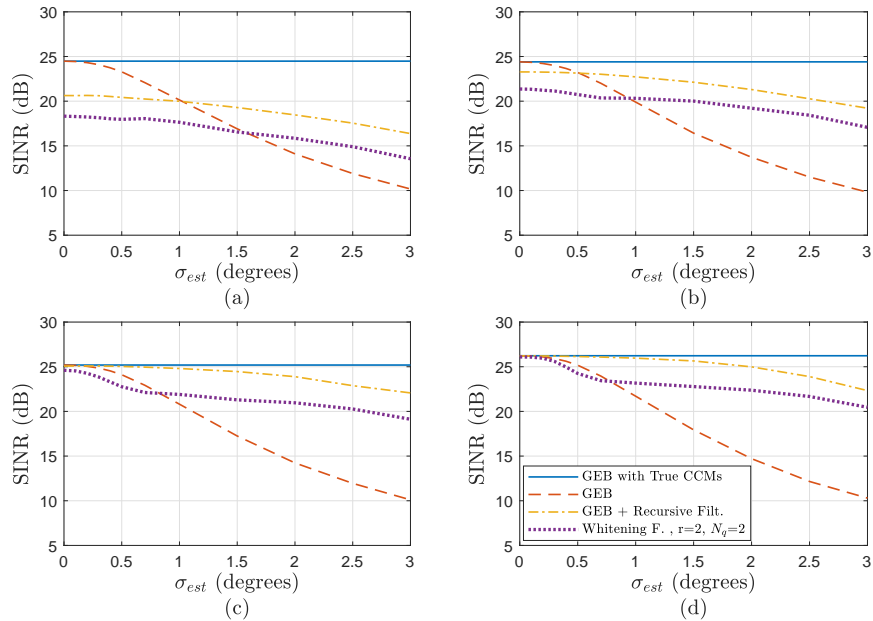}}
\caption[caption]{Output SINR performances of different methods versus $\sigma_{est}$; for group 2, $D_2=2+2=4$ RF chains, SZF, $\beta=0.9$. (a) $\alpha=0.9$. (b) $\alpha=0.99$. (c) $\alpha=0.999$. (d) $\alpha=0.9999$.}
\label{fig_SINR_sigmaest_2}
\end{figure}

Figure \ref{fig_SINR_sigmaest_1} is repeated for group 2 and results are shown in Figure \ref{fig_SINR_sigmaest_2}, using $d_1^{(2)}=2$ and $d_2^{(2)}=2$, therefore $D_2=4$. Spatial zero forcing is applied, therefore Monte-Carlo trials are performed. It is seen that performance decrease is limited with 4 dB SINR for GEB, which shows that the spatial zero forcing is successful in intra-beam multiplexing. Whitening Filter experiences a further decrease in SINR due to the fact that it is derived upon the assumption that CCM of an MPC can be approximated as a rank-one matrix. However, it is using further dimensions now since $d_m^{(g)}>1$ in this case.

\begin{figure}[h]
\centerline{\includegraphics[width=0.70\textwidth]{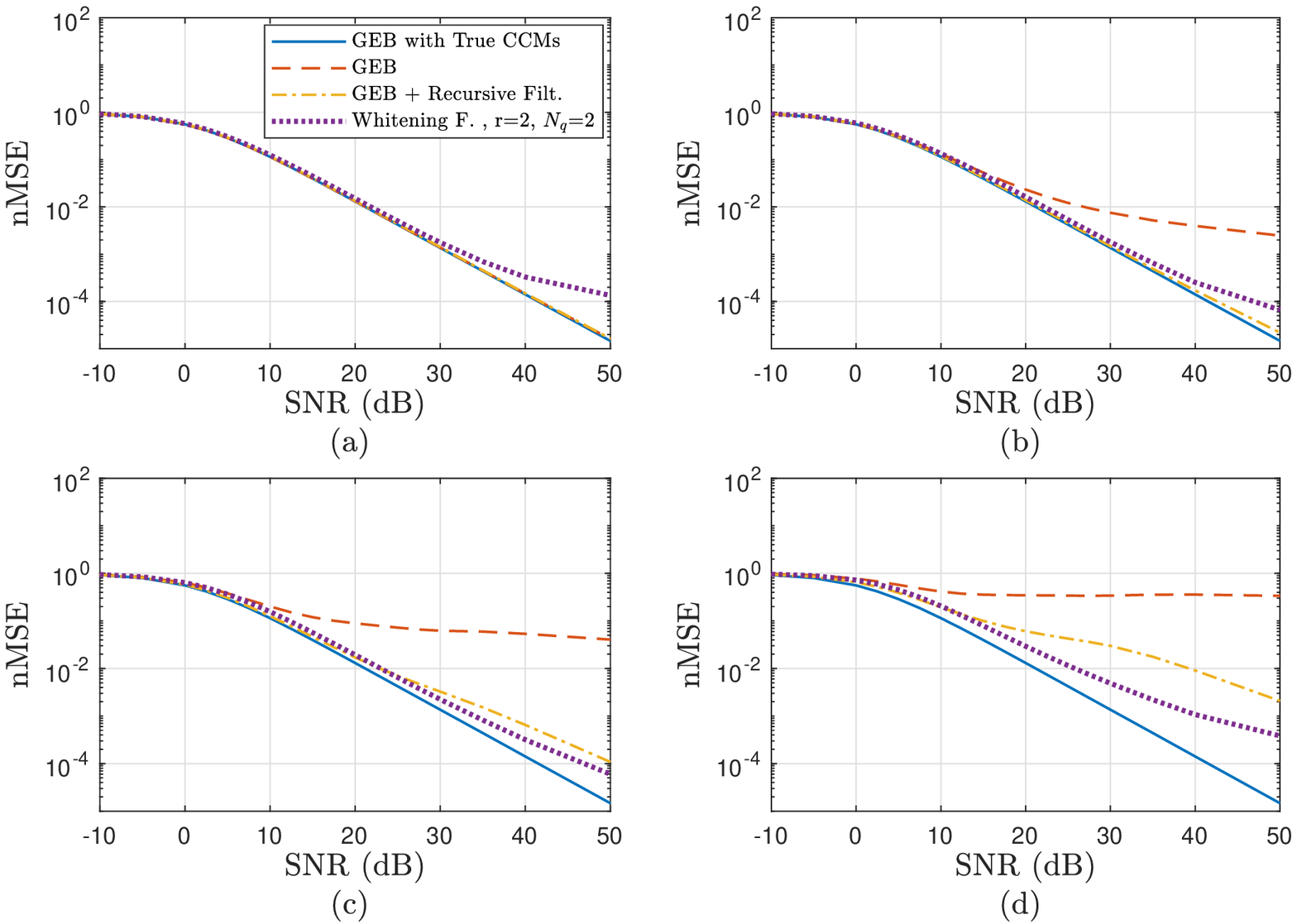}}
\caption[caption]{Channel acquisition performances of different methods versus input SNR; for training length $T=6$, $\alpha=0.999$, $\beta=0.9$. (a) $\sigma_{est}=0.1$. (b) $\sigma_{est}=0.5$. (c) $\sigma_{est}=1$. (d) $\sigma_{est}=2$.}
\label{fig_nMSE_SNR}
\end{figure}

\begin{figure}[h]
\centerline{\includegraphics[width=0.70\textwidth]{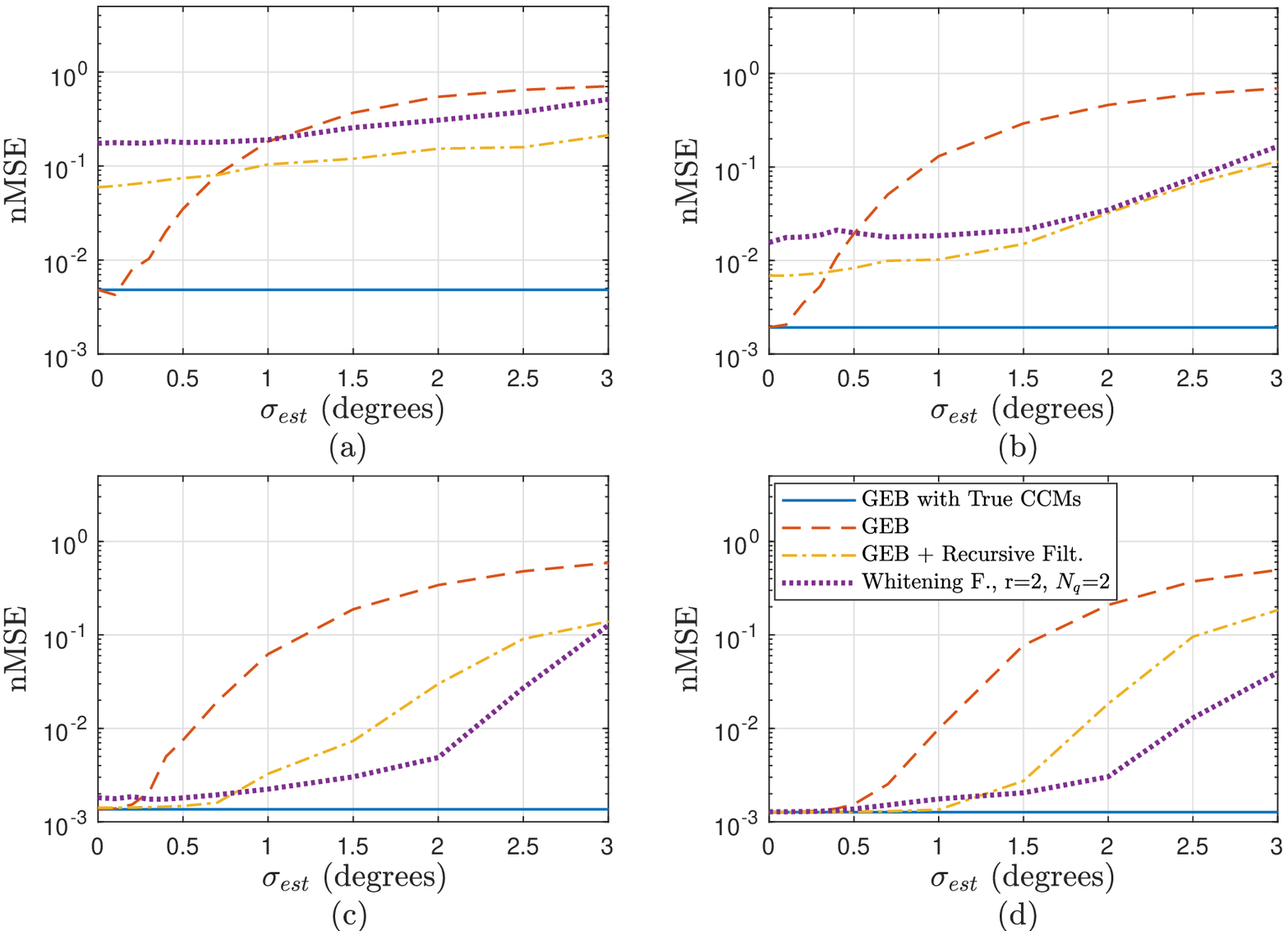}}
\caption[caption]{Channel acquisition performances of different methods versus $\sigma_{est}$; for training length $T=6$, $\beta=0.9$. (a) $\alpha=0.9$. (b) $\alpha=0.99$. (c) $\alpha=0.999$. (d) $\alpha=0.9999$.}
\label{fig_nMSE_sigmaest}
\end{figure}

In Figures \ref{fig_nMSE_SNR} and \ref{fig_nMSE_sigmaest}, instantaneous channel is not assumed to be perfect ($\hat{\mathbf{H}}_{eff,l}^{(g)} \neq \mathbf{H}_{eff,l}^{(g)}$), and reduced-rank approximated MMSE estimator is implemented. Using Monte-Carlo trials, normalized MSE is plotted against SNR and standard deviation of AoA estimation error $\sigma_{est}$. The results comfirm the comments made for Figure \ref{fig_SINR_sigmaest_1}.

%\section{Conclusion}
%The conclusion goes here.
\section{Conclusion}
In this paper, we proposed adaptive construction methods for statistical analog beamformer considering a time-varying, SC, wideband massive MIMO channel in mm-wave range. Proposed methods are shown to be robust against angular estimation errors, and very efficient in terms of computational complexity. Power profile estimation in angle-delay domain, user grouping, and resolution constraints of phase shifters and amplifiers in analog beamformer stage can be subjects of future studies.

 \appendix[Hadamard Factorization of CCMs and Angular Patching]\label{sechadfac}
The array response vector $\mathbf{u}(\phi)$ is dependent on azimuth angle $\phi$ through $\sin(\cdot)$ function. This fact is obstructive when trying to do derivations on CCMs, which are dependent on $\mathbf{u}(\phi)$. When the transformation from the azimuth angle $\phi$ to the phase $\theta$
\begin{equation}\label{transformation}
\theta \triangleq \pi \sin(\phi) \end{equation}
is performed, there exists a power profile $\rho_{\theta,l}^{(g)}(\theta)$ in phase domain, corresponding to $\rho_{\phi,l}^{(g)}(\phi)$, which satisfies
\begin{equation} \label{R_gen_int2}
\mathbf{R}_l^{(g)} = \int_{-\pi}^{\pi} {\rho}_{\theta,l}^{(g)}(\theta) \mathbf{q}(\theta) {\mathbf{q}(\theta)}^{H} d\theta ,
\end{equation} similarly to \eqref{R_gen_int} where $\mathbf{q}(\theta) \triangleq \frac{1}{\sqrt{N}}\left[1 \,\, e^{j \theta} \dots e^{j(N-1) \theta} \right]^{T}$.

Consider a power distribution ${\rho}_{\phi,l}^{(g)}(\phi)$ along the angular domain restricted between $\phi_1=\mu_{\phi,l}^{(g)}-\sigma_{\phi,l}^{(g)}/2$ and $\phi_2=\mu_{\phi,l}^{(g)}+\sigma_{\phi,l}^{(g)}/2$. Then, corresponding ${\rho}_{\theta,l}^{(g)}(\theta)$ is restricted between $\theta_1=\pi \sin(\phi_1)$ and $\theta_2=\pi \sin(\phi_2)$. Center angle $\mu_{\theta,l}^{(g)}$ and width $\sigma_{\theta,l}^{(g)}$ are defined according to $\theta_1$ and $\theta_2$. Consider  
\begin{equation} \label{patched_power_profile}
{\rho}_{\theta,l}^{(g)}(\theta) = \left(\sigma_{\theta,l}^{(g)}\right)^{-1} \text{rect}\left(\frac{\theta-\mu_{\theta,l}^{(g)}}{\sigma_{\theta,l}^{(g)}}\right),
\end{equation}
where $\text{rect}(\theta)$ is well-known rectangular function, equal to one for $-1/2 \leq \theta \leq +1/2$ and zero otherwise. When the integral in (\ref{R_gen_int2}) is computed for each entry of $\mathbf{R}_l^{(g)}$, it is found that 
\begin{equation} \label{eqhadamardfac}
\mathbf{R}_l^{(g)} = \left(\mathbf{q}(\mu_{\theta,l}^{(g)}) \mathbf{q}^{H}(\mu_{\theta,l}^{(g)}) \right) \odot \mathbf{D}(\sigma_{\theta,l}^{(g)})
\end{equation}
where $ \left(\mathbf{D}(\theta) \right)_{m,n} \triangleq \text{sinc}{\left((m-n)\frac{\theta}{2\pi}\right)} $, and $\odot$ is the Hadamard product operator. The CCM $\mathbf{R}_l^{(g)}$ is \emph{Hadamard-factorized} so that center angle is governed by a rank-1 matrix $\mathbf{q}(\mu_{\theta,l}^{(g)}) \mathbf{q}^{H}(\mu_{\theta,l}^{(g)})$, and angular spread is governed by a Toeplitz matrix $\mathbf{D}(\sigma_{\theta,l}^{(g)})$. 

If the angular domain is separated into parts by defining angular patches of width $\frac{2\pi}{N}$ about center angles $k \frac{2\pi}{N}$ for $k=0,1,\dots , N-1 $, the power profile can be approximated as 
\begin{equation} \label{eqangularpatching}
\rho_{\theta,l}^{(g)} (\theta) \cong \sum_{k=0}^{N-1} p_k \left(\dfrac{2\pi}{N}\right)^{-1} \text{rect}\left(
\dfrac{\theta-k\frac{2\pi}{N}}{\frac{2\pi}{N}}
\right),
\end{equation}
where power levels $p_k$ are selected by awarding equal powers to the patches in which occupancy from $\rho_{\theta,l}^{(g)}(\theta)$ is found; i.e.,
\begin{equation}
p_k=\left\lbrace \begin{matrix}  a , & \mu_{\theta} + \frac{\sigma_{\theta}}{2} > k\frac{2\pi}{N}-\frac{\pi}{N} \, \text{ and } \, \mu_{\theta} - \frac{\sigma_{\theta}}{2} < k\frac{2\pi}{N}+\frac{\pi}{N} &
\\
 0, & \text{otherwise} & \end{matrix} \right. 
\end{equation} 
for $k=0,\dots , N-1$, where $a$ is selected to satisfy total power equality as 
\begin{equation}
\sum_{k=0}^{N-1} p_k = \int_{-\pi}^{\pi} \rho_{\theta,l}^{(g)}(\theta) d\theta.
\end{equation}
Then, use (\ref{eqhadamardfac}) for each cell and sum them up to obtain
\begin{equation}\label{generalpatching}
\mathbf{R}_l^{(g)} = \left(\mathbf{Q} \mathbf{P}_l^{(g)} \mathbf{Q}^{H} \right) \odot \mathbf{D},
\end{equation}
where $\mathbf{Q} \triangleq \left[ \begin{matrix} & \mathbf{q}(0) & \mathbf{q}(\frac{2 \pi}{N}) & \cdots & \mathbf{q}((N-1) \frac{2 \pi}{N}) & \end{matrix} \right]$ is the DFT matrix of size $N$, and $\mathbf{P}_l^{(g)} \triangleq \text{diag}\left[\left\lbrace p_k\right\rbrace_{k=0}^{N-1}\right]$. For the sake of simplicity, $\mathbf{D}\triangleq\mathbf{D}(\frac{2 \pi}{N})$ and it is constructed as
\begin{equation}
\left(\mathbf{D}\right)_{m,n} = \text{sinc}\left(\frac{m-n}{N} \right).
\end{equation} 
Note that the matrices $\mathbf{Q}$ and $\mathbf{D}$ are constant matrices and only $\mathbf{P}_l^{(g)}$ is specific to $\mathbf{R}_l^{(g)}$. Accordingly,
\begin{equation}
\mathbf{R}_y \cong \left(\mathbf{Q} \mathbf{P}_y \mathbf{Q}^{H} \right) \odot \mathbf{D},
\end{equation}
where
\begin{equation}
\mathbf{P}_y \triangleq \sum_{g=1}^G K_{g}E_s^{(g)} \sum_{l=0}^{L-1} \mathbf{P}_{l}^{(g)} + N_0 \mathbf{I}_N.
\end{equation}

Let the vector $\mathbf{d}_n(\sigma_{\theta,l}^{(g)})$ be the multiplication of the square root of the $n$\textsuperscript{th} most dominant eigenvalue and the $n$\textsuperscript{th} eigenvector of the matrix $\mathbf{D}(\sigma_{\theta,l}^{(g)})$ as
\begin{equation} \label{eigenvectorsofD1}
\mathbf{D}(\sigma_{\theta,l}^{(g)}) = \sum_{n=1}^{N} \mathbf{d}_n(\sigma_{\theta,l}^{(g)}) \mathbf{d}_n^{H}(\sigma_{\theta,l}^{(g)}).
\end{equation}
Then, except a scalar multiplier, $n$\textsuperscript{th} most dominant eigenvector of $\mathbf{R}_l^{(g)}$ with $\rho_{\theta,l}^{(g)}(\theta)$ as in \eqref{patched_power_profile} is \cite{kurtmsc} ;
\begin{equation} \label{gevecRlg}
\mathbf{w}_{n,l}^{(g)} \triangleq \mathbf{q}(\mu_{\theta,l}^{(g)}) \odot \mathbf{d}_n(\sigma_{\theta,l}^{(g)}).
\end{equation}

%\section*{Acknowledgment}

\ifCLASSOPTIONcaptionsoff
  \newpage
\fi

\end{document}